\RequirePackage{fix-cm}
\documentclass[twocolumn,epjc3]{svjour3}
\RequirePackage{graphicx,subfigure}

\newcommand{\mm}{\ensuremath{\,\mathrm{mm}}}

\journalname{Eur. Phys. J. C}

\begin{document}
\title{Implementation of a local principal curves algorithm for neutrino
interaction reconstruction in a liquid argon volume} 
\titlerunning{Lpc reconstruction algorithm}

\author{J.J. Back\thanksref{e1,addr1}\and 
G.J. Barker\thanksref{addr1}\and 
S.B. Boyd\thanksref{addr1}\and 
J. Einbeck\thanksref{addr2}\and
M. Haigh\thanksref{addr1}\and
B. Morgan\thanksref{addr1}\and 
B. Oakley\thanksref{addr1}\and
Y.A. Ramachers\thanksref{addr1}\and
D. Roythorne\thanksref{addr1}}
\thankstext{e1}{e-mail: j.j.back@warwick.ac.uk}
\institute{Department of Physics, University of Warwick, Coventry CV4 7AL,
  UK\label{addr1}
\and
Department of Mathematical Sciences, Durham University, Durham DH1 3LE,
UK\label{addr2}}
\date{Received: 23 December 2013 / Accepted: 19 March 2014}
\maketitle
\begin{abstract}
A local principal curve algorithm has been implemented in three 
dimensions for automated track and shower reconstruction of neutrino interactions 
in a liquid argon time projection chamber. We present details of the algorithm and
characterise its performance on simulated data sets. 
\end{abstract}

%
\section{Introduction}
\label{sec:intro}

Liquid argon time projection chambers (LAr-TPCs), that
are currently in development in various R\&D programmes in Europe, Japan
and the USA~\cite{RDStatus}, are acknowledged to be a detector
technology capable of meeting the physics requirements of a
next-generation neutrino oscillation experiment. They can provide
simultaneous tracking and calorimetry of particles from neutrino
interactions over a wide range of energies, with exquisite
millimetric granularity, as demonstrated by results from
ICARUS~\cite{ICARUS}. Despite this advantage, it has proven difficult
to achieve an automated software process that can fully reconstruct
neutrino interactions, which will contain a mixture of 
ionisation tracks as well as electromagnetic and hadronic showers, especially 
when the neutrino interaction point is not known beforehand. In this paper, we
describe a first application of using local principal curves~\cite{ETE05} to 
automatically reconstruct neutrino interactions using three-dimensional 
LAr-TPC data.

When a charged particle passes through a liquid argon medium it releases
a stream of ionisation charge which can be measured by a TPC
to provide a 3D trajectory in space. We can represent this data as a 
collection of ``hits'', each of which contain the spatial cell co-ordinate
information ($x,y,z$) as well as the charge or energy deposit $Q$.
The task of any reconstruction algorithm is to first obtain the hits
from the detector output, then group these hits into clusters in order to
identify the particles coming from the neutrino interaction, before extracting
physics parameters such as momentum or energy from the reconstructed particles.
Here, the first stage of the analysis chain, hit reconstruction, is assumed 
to have already taken place and the input is taken to be the complete set of hits
in three spatial dimensions. Our reconstruction algorithm
takes the collection of hits for each neutrino interaction (labelled as an event)
and forms clusters of associated hits in order to identify the particles.
The mathematics and logic behind the local principal curve procedure is described in
Sect.~\ref{sec:algo}, while a description of the simulation methods used to 
obtain samples of neutrino interaction events in liquid argon is given in Sect.~\ref{sec:sim}.
The performance of the reconstruction algorithm is discussed in Sects.~\ref{sec:mup}
and~\ref{sec:ep}, with details about using it for track-shower discrimination
provided in Sect.~\ref{sec:shower}. Finally, we summarise our findings in Sect.~\ref{sec:con}.

\section{The local principal curve (lpc) algorithm}
\label{sec:algo}

The key component of the method we are proposing is the mean shift procedure,
a versatile tool which is popular mainly in the computer vision 
community~\cite{ComMeer02}. In essence, the mean shift moves a point to the 
local mean of the data around this point.
For our case, the points are the positions $X_i$($x$, $y$, $z$) of all of the
hits, which are each scaled by their range, defined to be the difference between
the largest and smallest values of $X_i$
(though it is also possible not 
to scale at all, or to scale by other measures of spread such as the 
standard deviation). The local mean $m(u)$ for a set of $N$ hits is defined as
\begin{equation}
\label{eqn:locmean}
m(u) = \frac{\sum_{i=1}^N w_i(u) X_i}{\sum_{i=1}^N w_i(u)},
\end{equation}
where the weights $w_i(u)$, which determine the size and 
shape of the local neighbourhood at a chosen location $u$, are 
monotonically decreasing with increasing distance from $u$ to $X_i$.
A common choice of weights is the Gaussian density function
\begin{equation}
\label{eqn:weight}
w_i(u) = \frac{Q_i}{(2\pi)^{3/2}h^3} \exp\left\{-\frac{1}{2h^2} (X_i - u)^T (X_i - u) \right\},
\end{equation}
where $Q_i$ is the energy deposit for hit $i$ and $h$ is a constant bandwidth 
parameter that steers the size of the local neighbourhood. The weights play 
the role of ``kernel'' functions and 
can, if desired, be replaced by other functions such as a 
triangular-shaped or truncated probability density.
In our scenario, where the co-ordinates are all measured on the same 
scale, we keep the Gaussian form and use the same bandwidth parameter for all 
three directions. The value of $h$ can be selected through a coverage 
measure~\cite{ETE05}, though for our purposes there is 
not much reason for this, as roughly the same bandwidth, $h \sim 0.05$ after scaling, 
will be usable in a wide range of liquid argon detectors.
Note that the normalisation denominator $(2\pi)^{3/2}h^3$ can be left out
of the kernel function since it is a constant common factor for all hit points 
and has no effect on the properties of the principal curve.
\begin{figure*}[htb]
\centering
\includegraphics[width=0.75\textwidth]{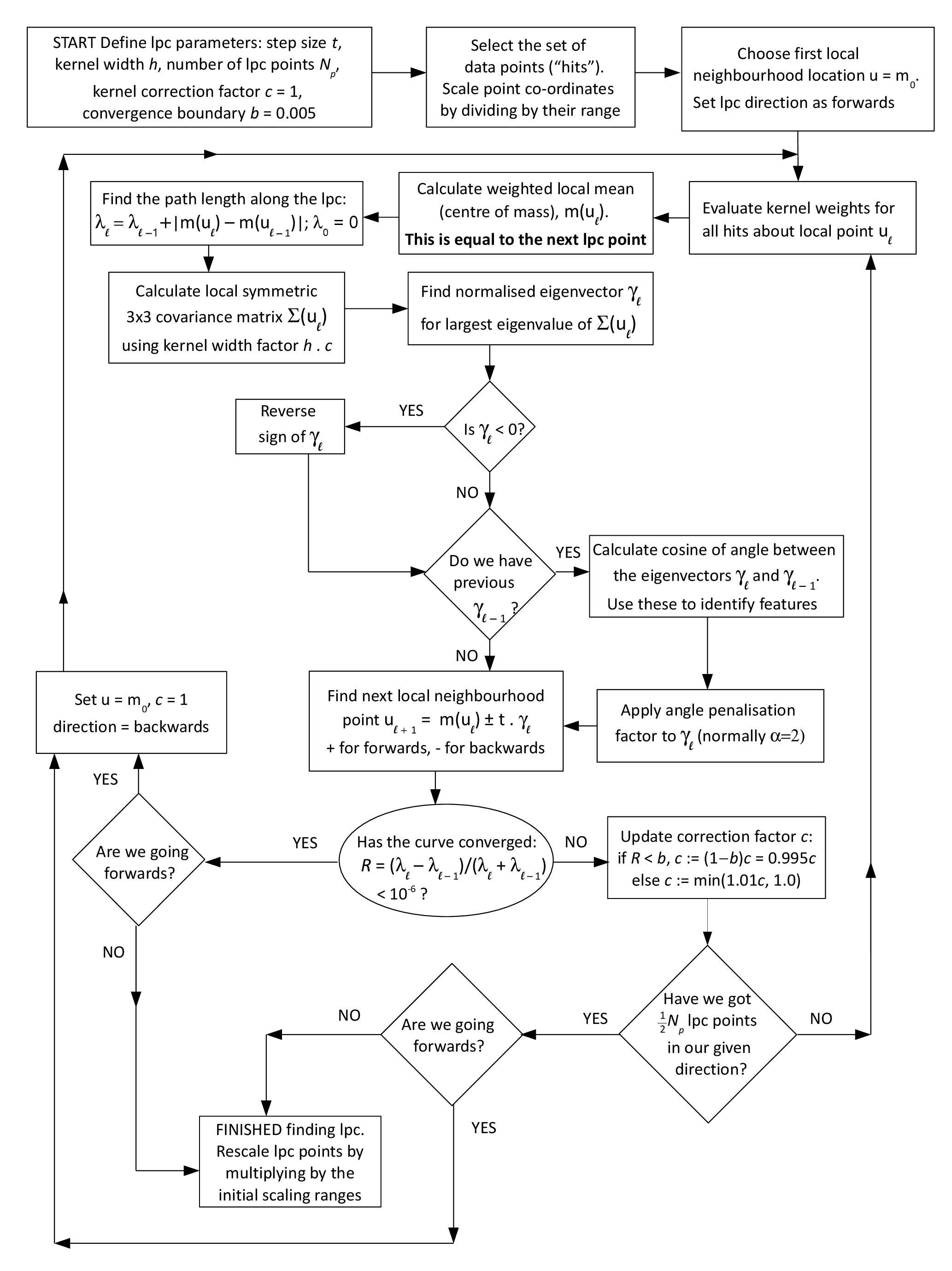}
\caption{Logic flow of the local principal curve (lpc) algorithm. 
The starting point is at the top left and the
arrows show the direction to the next action (rectangle) or 
decision (diamond or ellipse)}
\label{fig:algorithm}
\end{figure*}

From Eq.~\ref{eqn:locmean}, we define the \emph{mean shift} as 
\begin{equation}
\label{eqn:ms}
s(u) = m(u) - u = \frac{\sum_{i=1}^N w_i(u) (X_i-u)}{\sum_{i=1}^N w_i(u)}.   
\end{equation}
This quantity has many interesting properties~\cite{ComMeer02}, one of which being that 
$s(u) \propto \nabla \hat{f}(u)/ \hat{f}(u)$, where 
$\hat{f}(u)= \frac{1}{N}\sum_{i=1}^N w_i(u)$ is a density estimate of $f$ at $u$.
This implies that the mean shift is a vector pointing into a denser direction of 
the data space. When carried out iteratively, starting at $u = m_0$, one can
show~\cite{ComMeer02} that the series of local means
\begin{equation}
\label{eqn:itmeans}
m_{\ell+1} = m_{\ell}+ s(m_{\ell}), \quad \ell \ge 0,
\end{equation}
converges to a local mode $u_m$ of $\hat{f}(u)$ where $s(u_m)=0$.
This has the attractive property of being a clustering technique; a trajectory can
be formed by running the mean shift 
procedure on each data point $X_i$ iteratively until convergence is achieved.

Though the convergence towards a local mode of the density is an appealing 
property, it has the negative side effect of getting trapped at the local modes 
and will not move beyond them. Therefore, 
some modification of Eq.~\ref{eqn:itmeans} is needed which ensures 
that particle trajectories are pursued beyond local modes. The simple idea is to 
alternate the mean shift with a local principal component step~\cite{ETE05}. 
More specifically, let $\gamma(u)$ be the normalised eigenvector corresponding to 
the largest eigenvalue of the local symmetric $3 \times 3$ covariance matrix
\begin{equation}
\label{eqn:sigma}
\Sigma(u)= \frac{\sum_{i=1}^N w_i(u)(X_i-m(u))(X_i-m(u))^T}{\sum_{i=1}^N w_i(u)}.
\end{equation}
Starting from a given point $u = m_0$, we set $\ell=0$ and iterate between
\begin{enumerate}
\item computing the local centre of mass:
\begin{equation}
\label{eqn:mu}
m(u_{\ell}) \equiv u_{\ell} + s(u_{\ell});
\end{equation}
\item finding the next local neighbourhood location:
\begin{equation}
\label{eqn:u}
u_{\ell+1} = m(u_{\ell}) + t \times \gamma_{\ell},
\end{equation}
\end{enumerate}
where $t$ is a given step size (of the same order as $h$) and 
$\gamma_{\ell} \equiv \gamma(u_{\ell})$. The local principal curve is then 
defined as the series of local centres of mass $m(u_{\ell})$.
In our case, the starting point $u = m_0$ is chosen to be the position of the 
nearest hit to the energy-weighted centroid of all of the hits. 
Alternatively, $m_0$ can be set either at random from the $X_i$ points,
set by hand, or be chosen to be a local density mode using an initial mean 
shift procedure as outlined in Refs.~\cite{ETE05,Zayed11}.
The above iteration is repeated until either
the required number of lpc points ($N_p$) is obtained, 
or the path length along the local curve is no longer increasing (convergence).

As we will see later, the angle $\phi$ between the 
normalised eigenvector $\gamma_{\ell}$ and the preceeding 
eigenvector $\gamma_{\ell-1}$ can be used to infer the presence
of feature points, corresponding to drops in the angle profile along the principal curve,
which provide evidence for particle decays or possible 
interactions between particles.

In order to provide inertia for reducing the chance of the local principal
curve deviating too much from the general direction of neighbouring points,  
$\gamma_{\ell}$ is multiplied by an angle penalisation 
term $a = |\rm{cos}\phi|^{\alpha}$, where $\alpha$ is usually set
to 2, when the next local neighbourhood location is found using Eq.~\ref{eqn:u}:
\begin{equation}
\label{eqn:gamma}
\gamma_{\ell} := a \gamma_{\ell} + (1 - a) \gamma_{\ell - 1}.
\end{equation}
A technicality to be mentioned is that, for a given $\Sigma(u)$, the first 
eigenvector may equally well be $-\gamma{(u)}$ as well as $\gamma{(u)}$, so 
for each $\ell \ge 1$, one needs to check whether 
$\rm{cos}\phi > 0$ and set $\gamma_{\ell}:=-\gamma_{\ell}$ otherwise~\cite{ETE05}.

Based on asymptotic considerations~\cite{Zayed11}, it can be shown
that the sequence of lpc points $m(u_{\ell})$ converges to a point $u_b$ close to 
the boundary of the cloud of data points with the property $f(u_b)= h |\nabla \hat{f}(u_b)|$.
In practical terms, convergence is reached when the cumulative path length difference
between neighbouring local curve points, 
divided by their sum, is below a chosen threshold typically set at $10^{-6}$:
\begin{equation}
\label{eqn:R}
R = \frac{\lambda_{\ell} - \lambda_{\ell - 1}}{\lambda_{\ell} + \lambda_{\ell - 1}} < 10^{-6},
\end{equation}
where $\lambda_{\ell} = \lambda_{\ell - 1} + |m(u_{\ell}) - m(u_{\ell - 1})|$ and $\lambda_0 = 0$.
In order to pick up features that may be present in the tails of the point cloud,
convergence is delayed by multiplying the kernel bandwidth $h$ with a correction
factor $c$ if $R$ is below a certain boundary limit $b$, which is
typically set to be $5\times10^{-3}$ (which must be
larger than the threshold for $R$ defined in Eq.~\ref{eqn:R}).
Initially, $c$ is set to unity, but when $R < b$, 
$c$ is reduced by the factor $(1-b)$, and $h$ in Eq.~\ref{eqn:weight} needs to be 
replaced by $h \times c$. If $R \geq b$, then $c$ is increased by 1\% but must not exceed
unity. After convergence, or after we have obtained $\frac{1}{2}N_p$ lpc points,
the algorithm has to be re-started with $u = m_0$
in order to cover the other side of the data cloud, where the next neighbourhood
location is defined as
\begin{equation}
\label{eqn:uback}
u_{\ell+1}=  m(u_{\ell}) - t \times \gamma_{\ell},
\end{equation}
and the lpc algorithm continues as usual.

Note that the cumulative path lengths $\lambda_{\ell}$ form a discrete parameterisation of the
principal curve, which can be refined via a cubic spline interpolation 
towards a continuous parametrisation if necessary~\cite{EEH10}. This can be useful since 
it allows the option to plot, and regress, physical quantities such as the amount of 
deposited energy as a function of distance along the particle trajectory covered by the local 
principal curve.

Figure~\ref{fig:algorithm} summarises the complete logic flow of the lpc
algorithm. Once the lpc points are found, they are scaled-up using the initial 
co-ordinate ranges of the hits. The important parameters with suggested (scaled) 
values are given in Table~\ref{tab:lpcparam}. 
These parameters are optimised to provide the best overall reconstruction performance 
for specific classes of neutrino interaction events described 
in Sects.~\ref{sec:mup} and~\ref{sec:ep}.
\begin{table}[!htb]
\caption{Default parameters for the lpc algorithm}
\centering
\begin{tabular}{ll}
\hline
Kernel bandwidth factor $h$ & 0.05 \\
Neighbourhood step size $t$ & 0.05 ($t \sim h$) \\
Number of lpc points $N_p$ & 100 to 200 \\
Initial kernel bandwidth correction $c$ & 1 \\
Angle penalisation factor $\alpha$ & 2 \\
Convergence boundary limit $b$ & 0.005 \\
Convergence criteria threshold $R$ & $10^{-6}$ \\
\hline
\end{tabular}
\label{tab:lpcparam}
\end{table}

\section{Simulation of neutrino interactions in liquid argon}
\label{sec:sim}

In order to test the reconstruction performance of the local principal curve algorithm for
neutrino interaction events, the Geant4 simulation toolkit~\cite{Geant2003} was used 
to implement a model of a LAr-TPC detector, defined to be a stainless steel cylinder 
with height and radius both equal to 10\,m centred at the origin (0,0,0)
and filled with liquid natural argon. Particles are tracked
through the detector volume with all electromagnetic and hadronic processes enabled.
The ``QGSP\_BIC\_HP'' physics list is used to model the hadronic interactions,
combining a quark-gluon string and binary cascade model with
high precision low-energy (below $20$\,MeV) neutron cross-section data.
The detector is divided into ``voxels'' with volumes equal 
to $(1 \times 1 \times 1)$\,\mm$^3$, and all primary and secondary
particles are tracked through these down to an energy of 10\,keV or until they leave 
the TPC volume. Energy deposits by charged particles passing through the voxels are
tallied into a map between the co-ordinates of the centres of each voxel
$(x,y,z)$ and the total deposited energy (charge) $Q$. To take into account
the effect of electron-ion recombination on the particle
stopping power in liquid argon, a quenching factor
is applied to all deposited energies using a modified form of Birks' law according
to results obtained from the ICARUS project~\cite{Quench}.
No attempt is made to model the detector readout system since this 
is highly experiment-specific.
The GENIE~\cite{Genie2010} package is used to simulate the primary
particles from muon-neutrino and electron-neutrino
interactions with a monoenergetic spectrum at
0.77\,GeV, which corresponds to the JPARC neutrino beam mean energy. The
neutrinos are directed in a beam along the $x$-axis through the centre (0,0,0) 
of the detector. In order to remove random hits from
secondary low-energy interactions, an initial filtering is applied to all of 
the hits using a density-based spatial clustering algorithm~\cite{DBScan1,DBScan2}.
Hits are required to be part of density-connected regions which contain
at least 10 hits, whereby the maximum allowed distance between a hit and 
its nearest neighbour is 2\,cm. Excluded hits are classified as ``noise'' and 
are removed from further processing.

\section{Charged--current quasi--elastic interactions: 
$\nu_\mu + n  \rightarrow \mu + p$}
\label{sec:mup}
\begin{figure}[htb]
\subfigure[]{
  \includegraphics[width=0.45\textwidth]{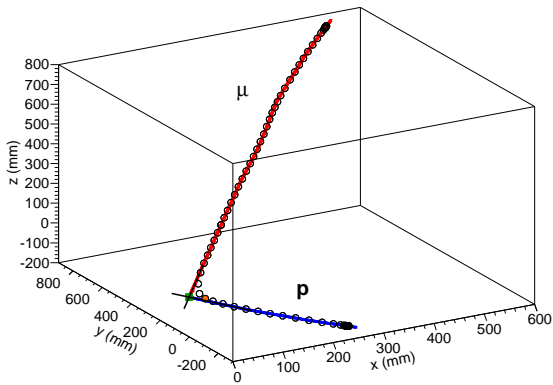}
}
\subfigure[]{
  \includegraphics[width=0.45\textwidth]{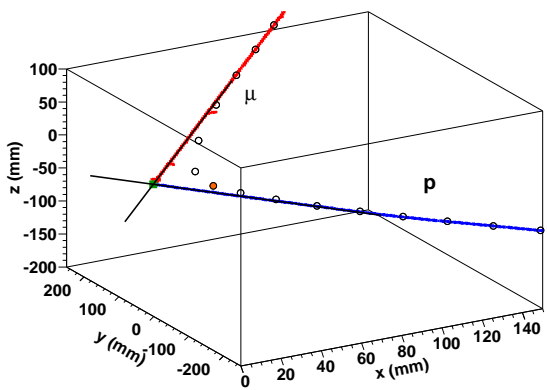}
}
\caption{An example lpc reconstruction of a muon-proton event showing
the hits associated to the muon (red) and proton (blue) tracks together 
with the calculated lpc points (open circles). Also shown are the 
two line segments used to find the position of the primary 
interaction vertex (green square). These lines are made from
hits on either side of the feature point of the principal curve, which 
has the largest $1-|\rm{cos}\phi|$ value, and is shown as an 
orange-filled circle. Plot (b) is a 
close-up view of the interaction region of plot (a)}
\label{fig:mupEvent}
\end{figure}

The suitability of using the local principal curve algorithm to 
reconstruct neutrino interactions can be demonstrated by its ability to
identify muon-neutrino charged-current quasi-elastic (CCQE) events, which 
have a simple two-track topology involving a short proton track
and a long muon track originating from a common primary vertex point,
with variable opening angle. Figure~\ref{fig:mupEvent} shows an example
reconstruction of a 770\,MeV muon-neutrino to muon-proton event, where it can be 
clearly seen that the calculated points on the curve follow the hits closely.
Along the middle portions of each track, the lpc points are roughly
equidistant from each other, which is an indication that the neighbouring
hits are essentially along a straight line. At the end of each track,
the lpc points begin to clump together as they approach convergence.

\begin{figure}
\centering
\includegraphics[width=0.45\textwidth]{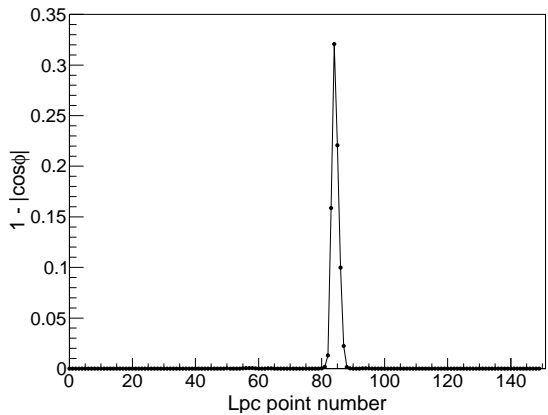}
\caption{Graph showing the quantity $1-|\rm{cos}\phi|$, for the example
muon-proton event shown in Fig~\ref{fig:mupEvent}, for points along the 
principal curve, where $\phi$ is the angle between the eigenvector 
$\gamma_{\ell}$ and the preceeding eigenvector $\gamma_{\ell-1}$ for 
lpc point $\ell$. The feature point is identified as the peak}
\label{fig:mupcosangle}
\end{figure}
\begin{figure}[htb]
\subfigure[]{
  \includegraphics[width=0.45\textwidth]{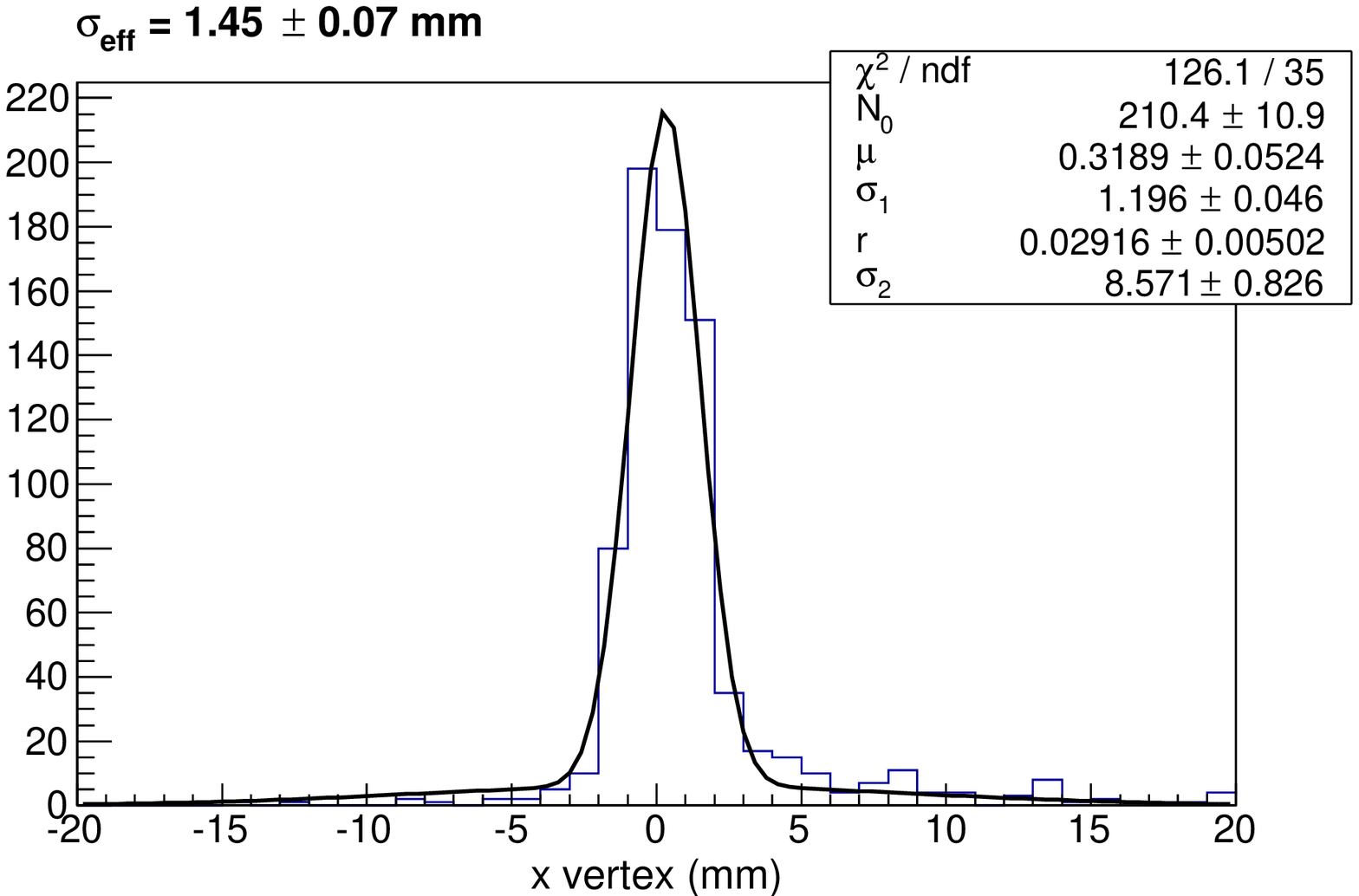}
}
\subfigure[]{
  \includegraphics[width=0.45\textwidth]{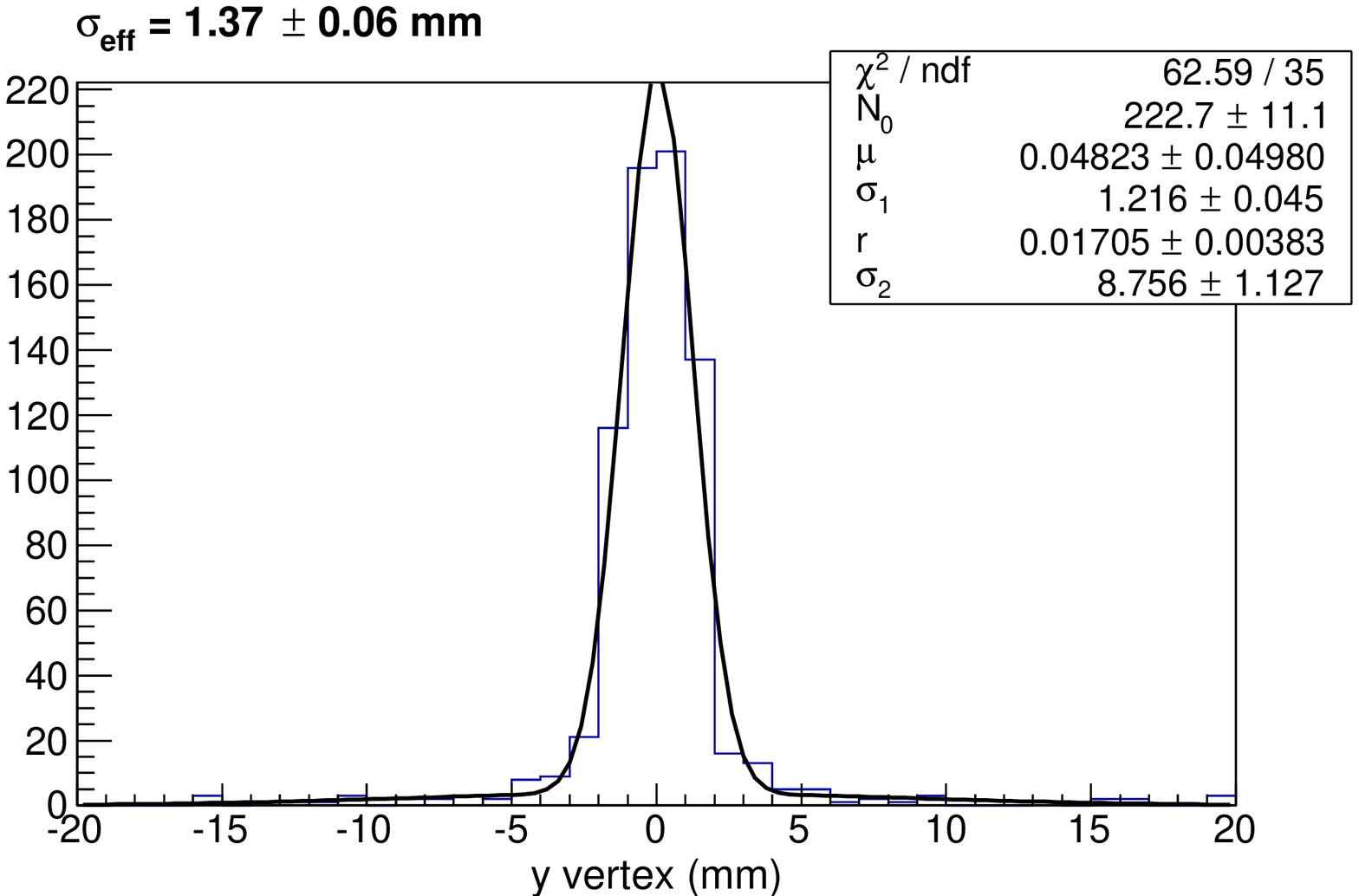}
}
\subfigure[]{
  \includegraphics[width=0.45\textwidth]{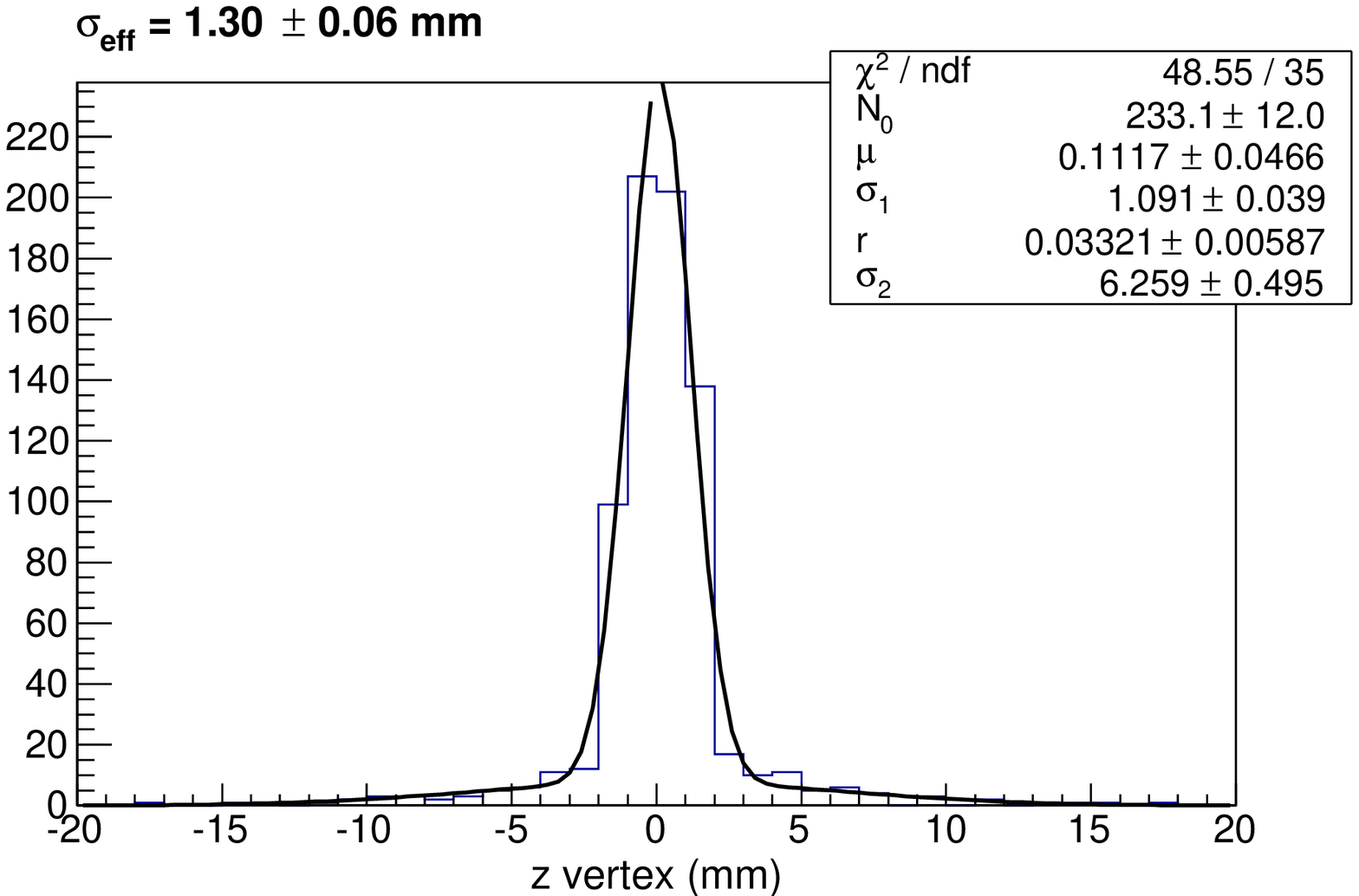}
}
\caption{Double Gaussian fits to the distributions of the 
reconstructed primary vertex position in $x$, $y$ and $z$ for 
muon-proton events that satisfy the proton range requirement.
Approximately 90\% of these events have a primary vertex found 
within 2\,cm from the true vertex position (0,0,0)}
\label{fig:mupvtx}
\end{figure}

Near the primary vertex position, the curve points transfer from the muon
track onto the proton track. During this transition, the angle $\phi$ 
between the eigenvectors of neighbouring lpc points increases.
As shown in Fig.~\ref{fig:mupcosangle}, plotting the distribution 
$1-|\rm{cos}\phi|$ as a function of lpc point number (or alternatively
as a function of the cumulative path length $\lambda_{\ell}$)
will produce a peak that will identify the specific
lpc \emph{feature point} $\ell_f$ which has the largest angle $\phi$,
and can be used to reconstruct the interaction vertex. First, we ignore
the two lpc points on either side of the feature point, since
the local curve is still rapidly changing direction, and only consider
the points numbered between $\ell_{f+2},\ell_{f+4}$ on one side and 
$\ell_{f-4},\ell_{f-2}$ on the other side. Additionally, any other feature
points (with lower $1-|\rm{cos}\phi|$ peaks) 
that may exist between $\ell_{f-4}$ and $\ell_{f+4}$ are ignored and
are considered to be just part of the original feature point $\ell_f$. For 
each range, straight lines are fitted to the hits that are closest to the 
lpc points. Each line is defined as a single point, taken to be the centroid
of the nearby hits, and a direction, which is chosen to minimise the 
sum of the perpendicular Euclidean distances of each hit to the line.
Next, an initial value of the vertex position, which will have
a typical resolution of approximately 1\,cm, is taken to be the point
of closest approach between the two straight line sections. The reconstruction
precision of the vertex location can be significantly improved by extending 
the straight lines towards the direction of the initial vertex point by adding
hits that are closest to a given line. These additional hits improve
the accuracy of the new centroid and direction of the two straight lines.
The vertex position is then taken to be the point of closest approach to 
these extended lines, leading to an improved resolution of approximately 1.5\,mm.
It is important to emphasise that the above method can only
reliably reconstruct two-prong vertices.

\begin{figure}[htb]
\includegraphics[width=0.5\textwidth]{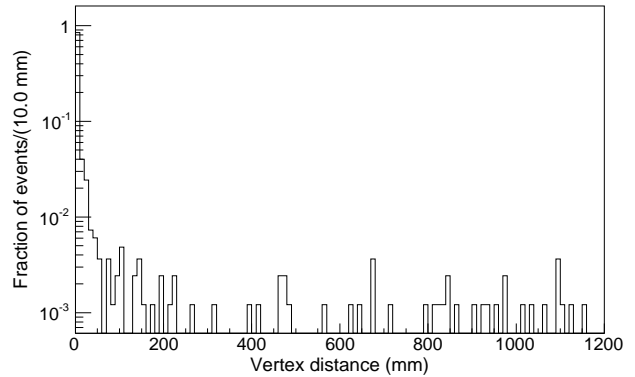}
\caption{The distribution of the distance of the 
reconstructed primary vertex position from the generated position (0,0,0) for
muon-proton events that pass the range selection on the proton track}
\label{fig:mupVtxdist}
\end{figure}

Figure~\ref{fig:mupvtx} shows the distributions of the primary vertex position
in $x$, $y$ and $z$ for a sample of 770\,MeV neutrino to muon-proton events when the
proton track has a minimum number of 25 hits (equal to a range 
of 2.5\,cm), which is roughly equivalent to an energy threshold of 
10\,MeV for a minimum ionising particle in liquid argon. Approximately
$15 \pm 1$\% of protons from a sample of $\nu_\mu + n \rightarrow \mu + p$ events 
will not satisfy this minimum range requirement.
The vertex distributions are fitted to
double Gaussian functions, which are defined to be the sum of two Gaussians
having the same mean $\mu$ but different widths $\sigma_1$ and $\sigma_2$,
with relative amplitude $r$.
The resolution of the vertex position in each co-ordinate direction is taken 
to be the effective width of the corresponding double Gaussian fit 
$\sigma_{\rm{eff}} = \sigma_1 + r\sigma_2$. As mentioned previously, the 
neutrino beam direction is defined to be along the $x$ axis, and so the muon and
proton tracks originating from the vertex will tend to have their largest 
momentum component along $x$. This has the effect of producing a very slight 
positive bias ($0.32 \pm 0.05$\,mm) for the determination of the $x$ position of 
the vertex point when using the above extended straight line method. Theoretically, this
bias could be reduced by having a smaller step size so that the principal
curve can get closer to the hits in the primary vertex region. In practice, this
does not significantly improve the overall vertexing performance, since reducing
the step size has the effect of increasing the occurance of fake 
secondary vertices (i.e. multiple  $1-|\rm{cos}\phi|$ peaks), since the algorithm 
becomes more suspectible to fluctuations in the hit point cloud.
As illustrated by Fig.~\ref{fig:mupVtxdist}, approximately 90\% of the events 
that pass the proton range selection have a primary vertex found within 2\,cm from 
the generated position (0,0,0).

\begin{figure}[htb]
\includegraphics[width=0.5\textwidth]{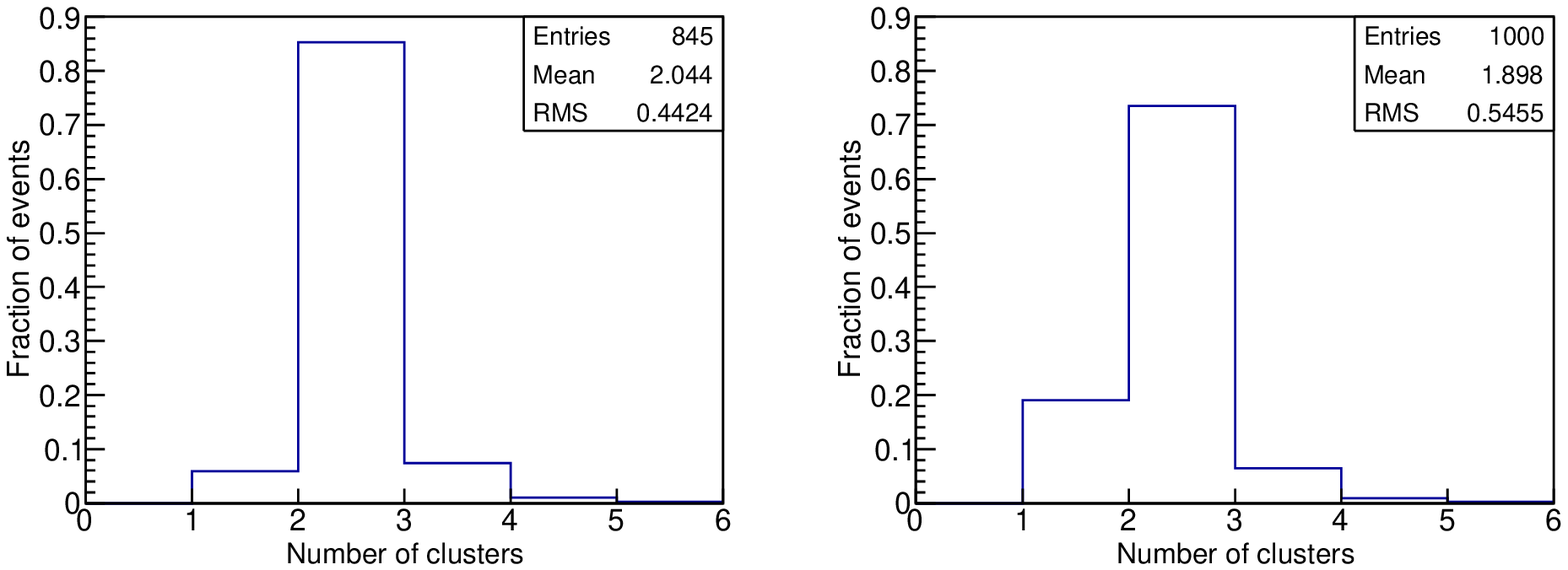}
\caption{The distribution of reconstructed clusters in a sample of
1000 muon-proton events for (left) events passing a range 
selection on the proton track and (right) with
no range requirement imposed}
\label{fig:mupNCl}
\end{figure}
\begin{figure}[htb]
\subfigure[] {
  \includegraphics[width=0.45\textwidth]{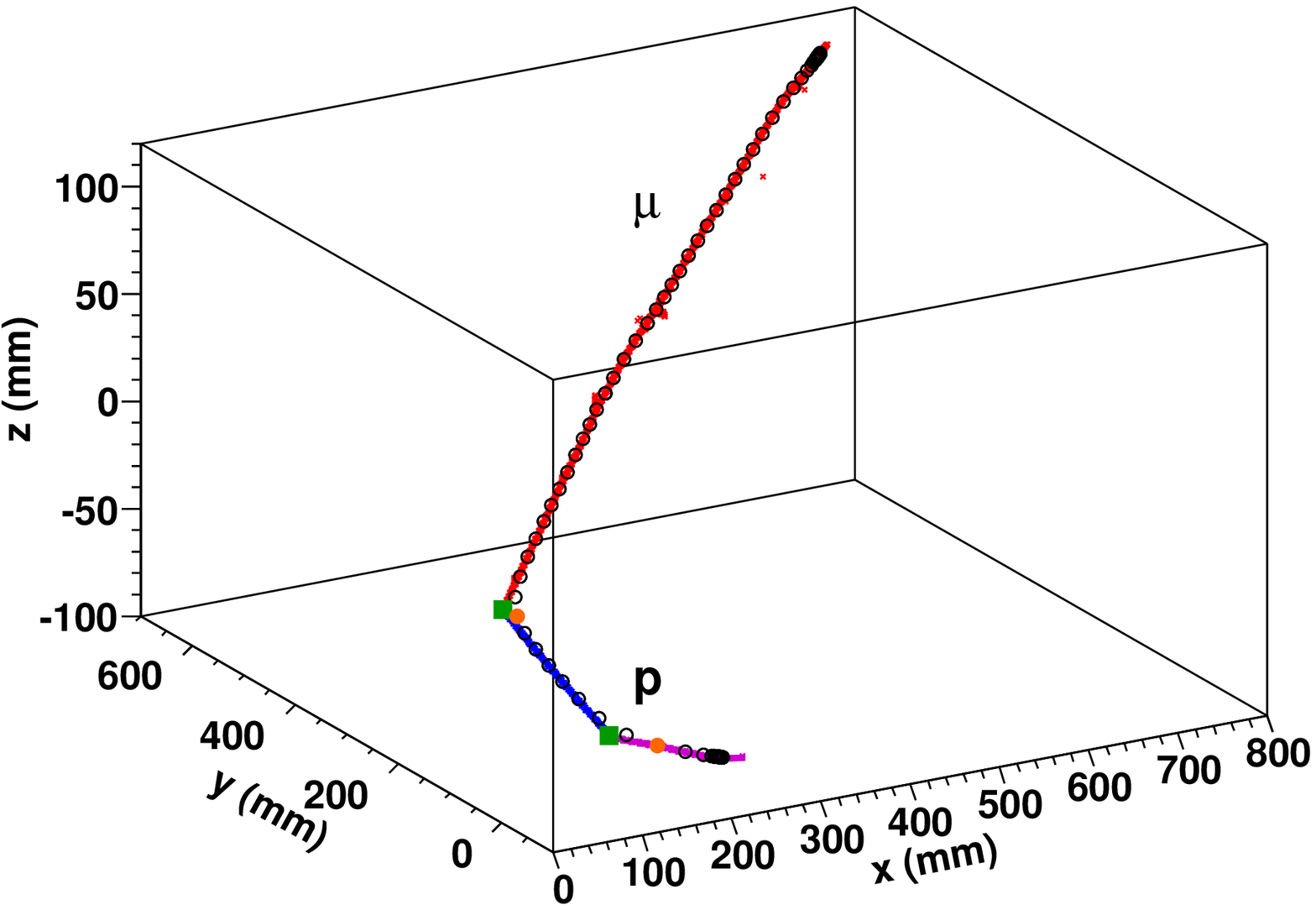}
}
\subfigure[] {
  \includegraphics[width=0.45\textwidth]{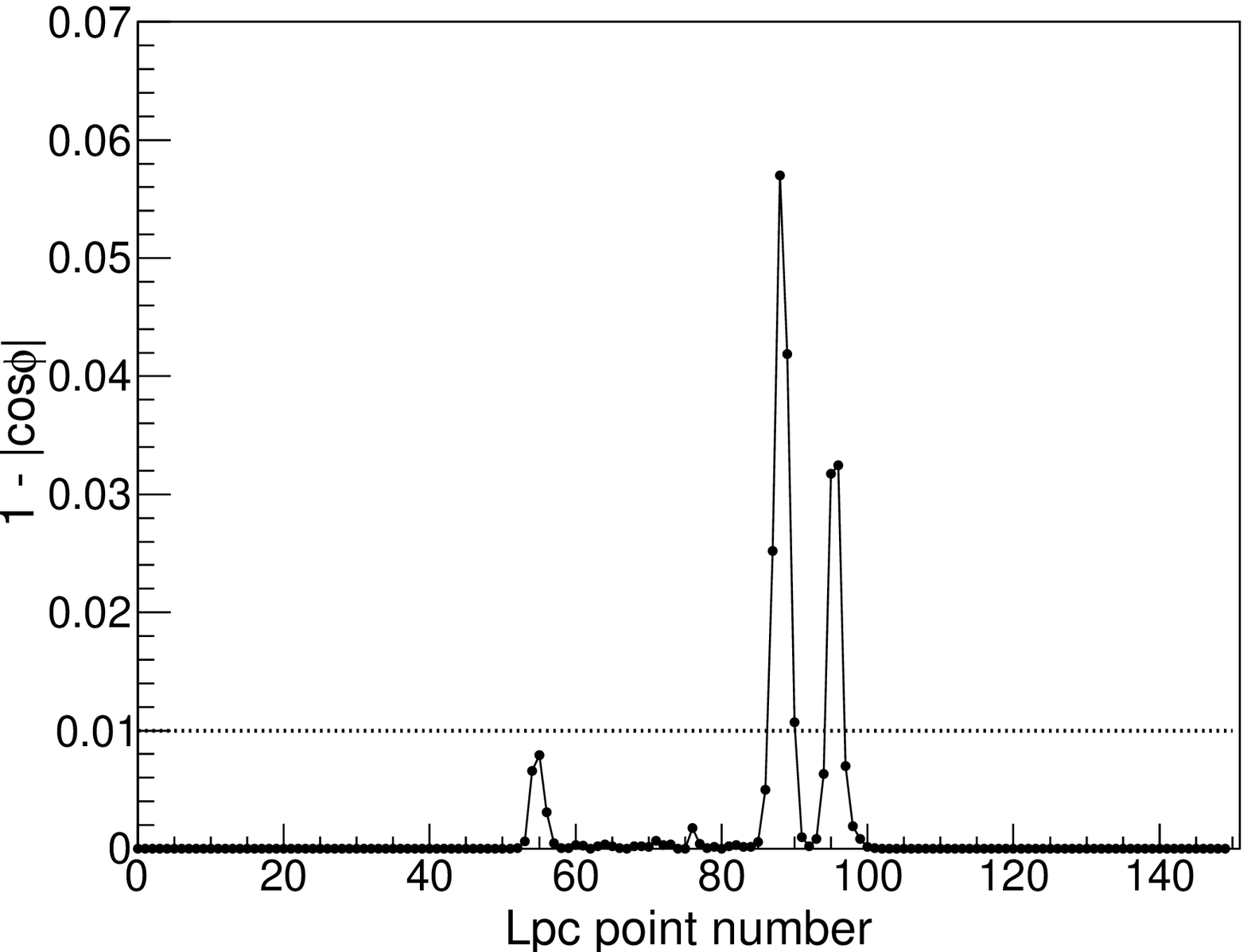}
}
\caption{An example lpc reconstruction of a muon-proton event where the proton
has been hard scattered. Image (a) shows the hits associated to the muon (red)
and proton (pre-scatter in blue, post-scatter in magenta) tracks.
Also shown are the calculated lpc points (open circles)
and feature points of the principal curve (orange-filled circles), together
with the reconstructed primary and secondary interaction vertices (green squares).
The feature points are those that correspond to the peaks in the $1-|\rm{cos}\phi|$
distribution shown in plot (b); these peaks are above the selection limit value of 0.01
represented by the dotted horizontal line}
\label{fig:mupInteract}
\end{figure}

Once a primary vertex has been found, it is then possible to assign
the hits on each extended straight line section to be the start of
separate \emph{clusters} for the proton and muon track. Further hits are added
to each cluster by continuing along the principal curve direction initially 
given by the straight line section and adding hits that are closest to the
remaining lpc points. Figure~\ref{fig:mupNCl} shows the number of clusters
found in muon-proton events with and without the proton range requirement. Most events
have just two clusters reconstructed, as expected, although about 6\% of the events
only have one cluster found, which occurs when the proton track has been missed,
i.e. the hits are just assigned to be the muon, and the vertex point is taken
to be just the start of the muon track.
More single cluster events (about 19\%) are reconstructed when no proton range selection
is imposed. For about 10\% of the events, more than two clusters are found, which 
occurs when two or more feature points are present with $1-|\rm{cos}\phi|$ values
above the threshold value of 0.01, which was chosen based on observations
of the typical heights of the feature peaks. Most of these additional 
secondary vertices, which are reconstructed using the same two-line method
described earlier, have genuine physics reasons: the proton or muon track 
scatters off a nucleus such as that shown in Fig.~\ref{fig:mupInteract}, or the 
muon decays to a low energy electron, producing a short two-prong stub at the 
end of the muon track. To reduce the chance of incorrectly finding secondary vertices,
neighbouring clusters that have principal axes within 20 degrees from each other are
merged and considered to be just one cluster, and the secondary vertex between them 
is removed. Varying this merging angle did not significantly improve the overall
reconstruction performance. When more than one vertex is found, the vertex with 
the lowest $x$ co-ordinate is chosen to be the primary vertex, since the neutrino 
beam is directed along the positive $x$ direction.
Note that the earlier primary vertex resolution plots shown in Fig.~\ref{fig:mupvtx} 
include events with secondary vertices found; only the vertex with the lowest 
$x$ value is included in the fitted distributions.

An important measure of the performance of this reconstruction algorithm is how well it
can correctly associate hits to each generated particle. The first
figure of merit is known as the average cluster efficiency $\epsilon_c$, which is 
equal to the number of reconstructed clusters which have the majority of the hits with
the correct particle type divided by the number of events. This quantity is strongly
correlated with the efficiency of finding a vertex, whereby the initial hit cloud
is broken up into the separate particle tracks (clusters).
The second figure of merit is the hit efficiency $\epsilon_h$ for each cluster, defined to
be the ratio of correct hits associated to the cluster compared to all hits 
produced by the original particle. Therefore, the overall efficiency of reconstructing a given 
particle is equal to the product of the cluster and hit efficiencies. Furthermore, the hit 
purity $\epsilon_p$ is defined to be the fraction of hits in a given reconstructed 
cluster that have the correct particle type.

The parameters of the principal curve defined in Table~\ref{tab:lpcparam} were optimised 
in order to give, on average, two clusters per muon-proton event, as well as providing
maximal cluster and hit efficiencies and purities for the reconstructed
muon and proton tracks. The only parameters that can significantly affect
the performance in this regard are the kernel width $h$, the 
step size $t$ and the number of lpc points $N_p$ (the other parameters are left unchanged).
Table~\ref{tab:mupeff} shows the results from this optimisation. It was found that 
variations to the kernel width and step size within the range
0.04 to 0.06 did not significantly affect the reconstruction performance,
and using between 100 and 300 lpc points also produced similar results.

\begin{table}[!htb]
\caption{Optimised performance of the lpc algorithm for 770\,MeV neutrino
to muon-proton events that satisfy the proton range requirement (845 out of an 
initial sample of 1000). Efficiencies and purities are averaged over all 
selected events}
\centering
\begin{tabular}{ll}
\hline
Quantity & Value \\
\hline
Lpc scaled kernel bandwidth $h$ & 0.056 \\
Lpc scaled step size $t$ & 0.040 \\
Number of lpc points & 150 \\
Fraction of events with no vertex found & $5.9 \pm 0.8$ \% \\
Muon cluster efficiency & $99.5 \pm 0.2$ \% \\
Muon hit efficiency & $93.7 \pm 0.8$ \% \\
Muon reconstruction efficiency & $93.2 \pm 0.9$ \% \\
Muon hit purity & $98.5 \pm 0.4$ \% \\
Proton cluster efficiency & $93.5 \pm 0.08$ \% \\
Proton hit efficiency & $91.7 \pm 0.9$ \% \\
Proton reconstruction efficiency & $85.8 \pm 1.2$ \% \\
Proton hit purity & $97.3 \pm 0.6$ \% \\
Vertex efficiency within $\pm 2$\,cm & $89.9 \pm 0.6$ \% \\
Vertex $x$ position resolution & $1.45 \pm 0.07$\,mm \\
Vertex $y$ position resolution & $1.37 \pm 0.06$\,mm \\
Vertex $z$ position resolution & $1.30 \pm 0.06$\,mm \\
\hline
\end{tabular}
\label{tab:mupeff}
\end{table}

\section{Shower and track discrimination}
\label{sec:shower}

\begin{figure}[htb]
\subfigure[]{
  \includegraphics[width=0.45\textwidth]{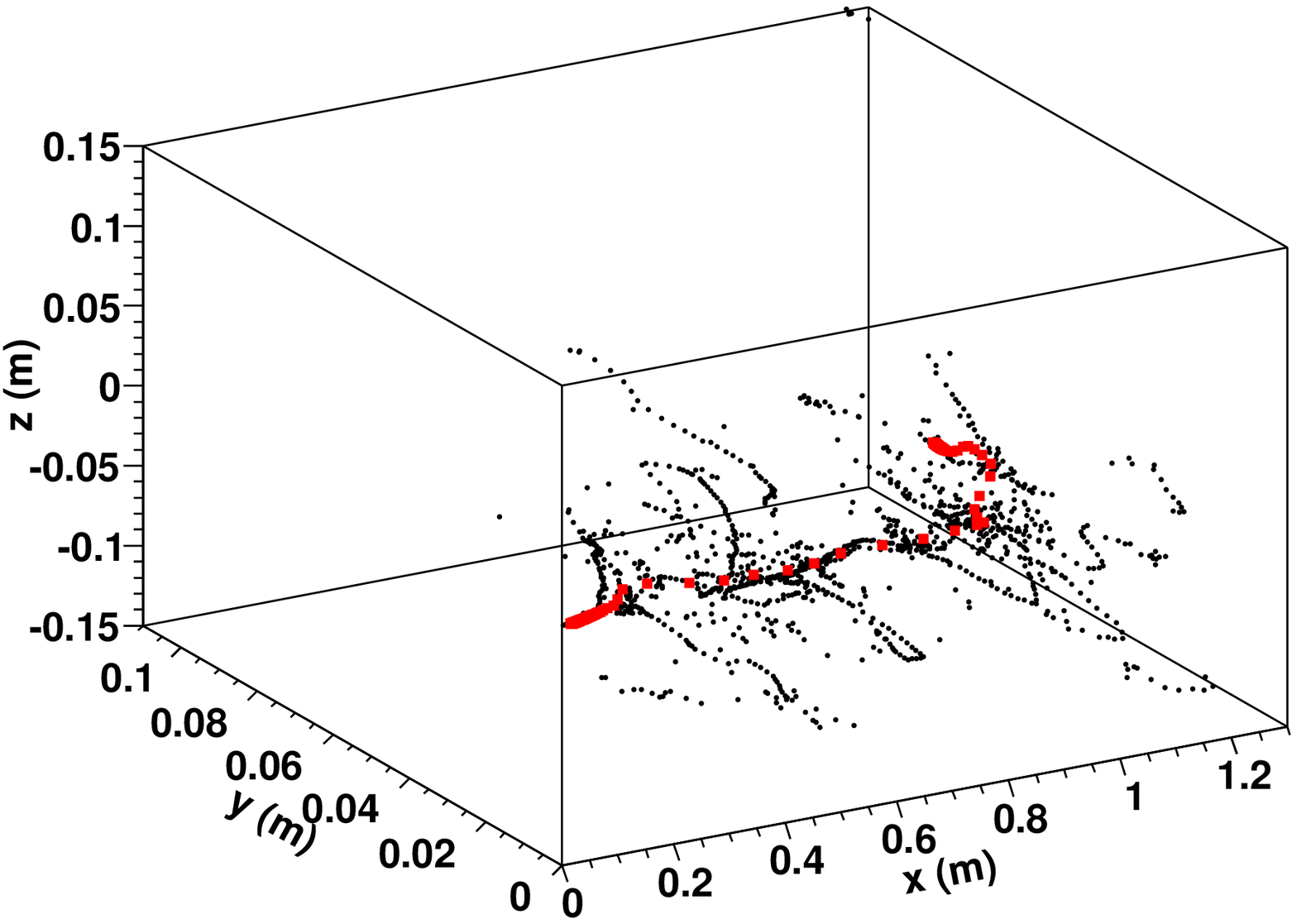}
}
\subfigure[]{
  \includegraphics[width=0.45\textwidth]{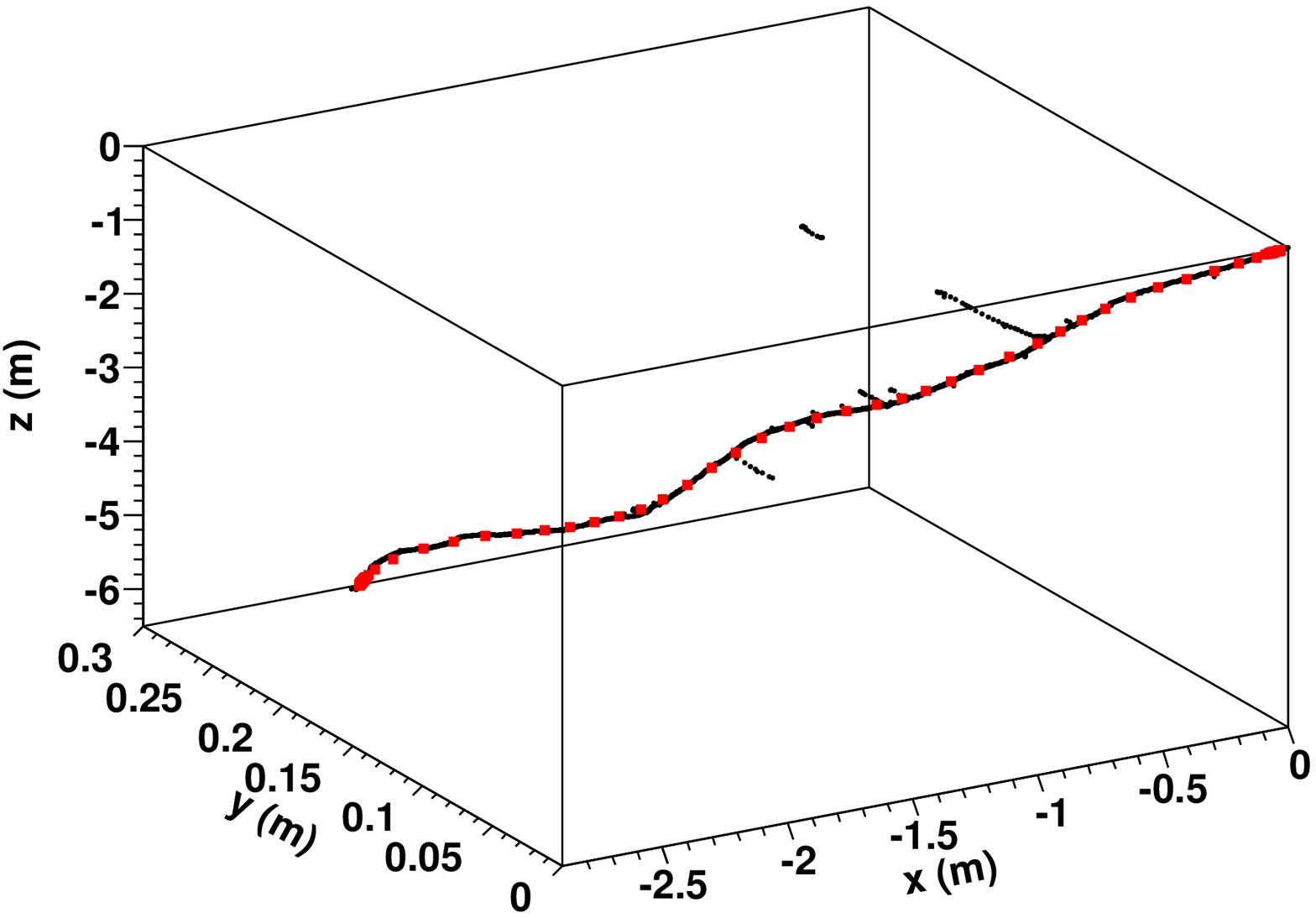}
}
\caption{Example principal curves (red squares) for 1.5\,GeV (a) electron shower and 
(b) muon track with delta electrons}
\label{fig:emu1_5GeV}
\end{figure}

We have seen that the principal curve algorithm has a very good
performance for reconstructing muon-neutrino CCQE
events. We next tested whether the algorithm can infer
the presence of electron-neutrinos via the interaction
$\nu_e + n \rightarrow e + p$, which means identifying
electron showers and proton tracks originating from a common vertex with
variable opening angle.
Before this can be attempted, we first need to implement a set of selection 
criteria that can tell us whether a cluster is either a shower or a track.
This can be achieved by looking at the differences between the transverse
and longitudinal extent of clusters. To avoid overcomplicating the
track-vs-shower analysis, only one principal curve is found per generated 
particle, and no vertex finding nor division into sub-clusters is performed.
As shown in Fig.~\ref{fig:emu1_5GeV}, an electron shower will generally 
produce a halo of hits that surround the principal axis of the point cloud, 
whereas a track will essentially be a continuous line of hits with slight 
changes in direction owing to the effects of multiple scattering. 
These topological differences can be quantified by looking at the 
Euclidean distance of each hit from its nearest calculated lpc point. 
These are known as \emph{residuals}, labelled as $\delta r$, and 
are a measure of the transverse extent of the hits in a cluster; in general, 
showers will have larger residuals.

\begin{figure}[htb]
\subfigure[]{
  \includegraphics[width=0.45\textwidth]{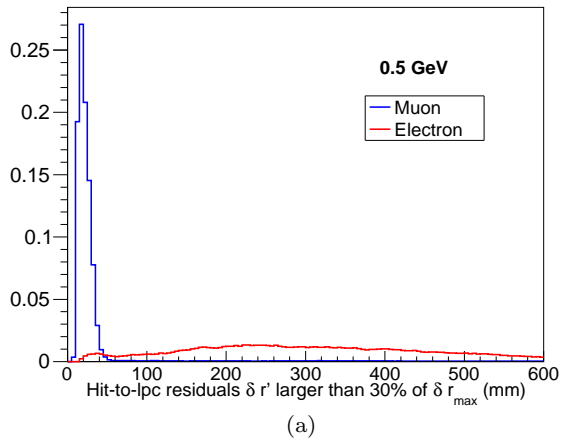}
}
\subfigure[]{
  \includegraphics[width=0.45\textwidth]{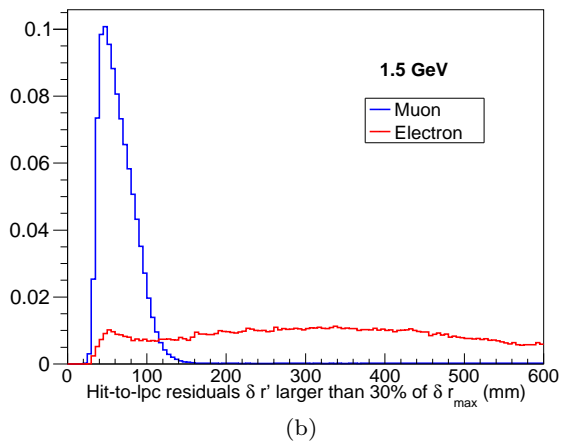}
}
\caption{Normalised distributions of the hit-to-lpc residuals $\delta r'$ for
(a) 0.5\,GeV and (b) 1.5\,GeV muon tracks and electron showers. Here, $\delta r'$
denotes residuals that are larger than 30\% of the value
of the maximum residual $\delta r_{\rm{max}}$}
\label{fig:resHist}
\end{figure}

\begin{figure}[htb]
\subfigure[]{
  \includegraphics[width=0.45\textwidth]{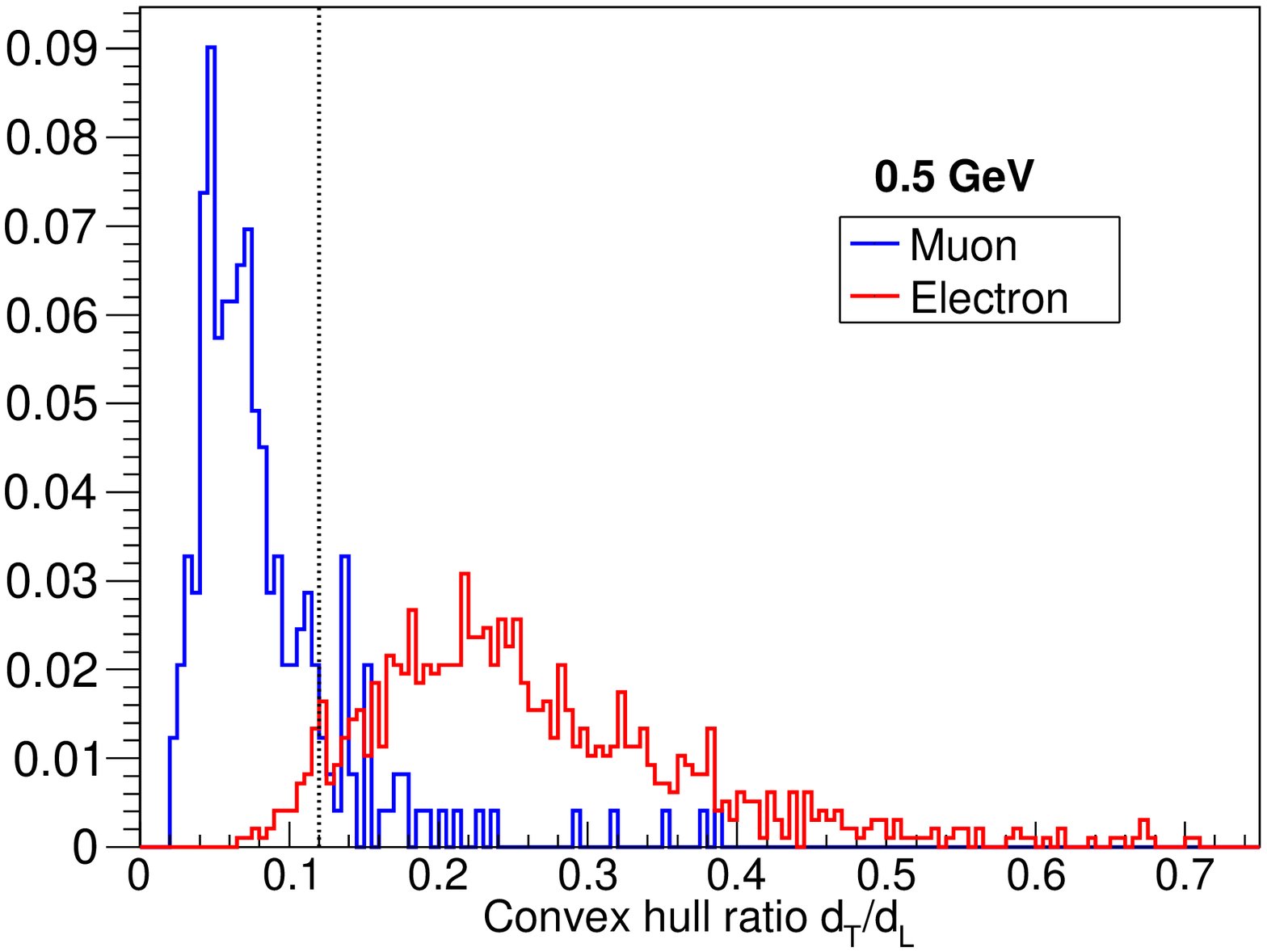}
}
\subfigure[]{
  \includegraphics[width=0.45\textwidth]{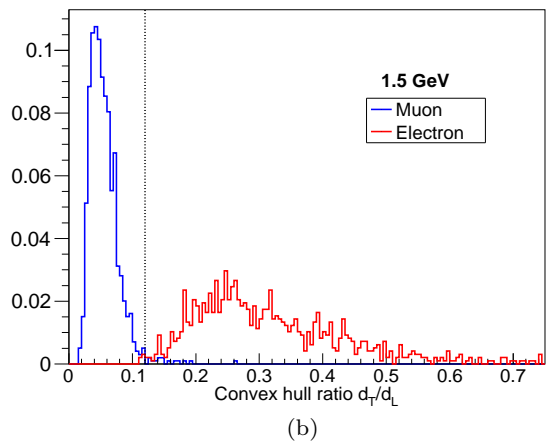}
}
\caption{Normalised distributions of the convex hull ratio for (a) 0.5\,GeV and (b)
1.5\,GeV muon tracks and electron showers that have at least 90\% of their hits with
$\delta r'$ at least equal to 2\,cm. The dotted vertical line represents the 
selection on the convex hull ratio; showers (tracks) have convex hull 
ratios above (at or below) 0.12}
\label{fig:convHull}
\end{figure}

Figure~\ref{fig:resHist} shows
the distribution of hit-to-principal-curve residuals
for 1,000-event samples of single particle, monoenergetic (0.5 and 1.5\,GeV) electrons
and muons. In order to enhance the differences between tracks and 
showers, very small residuals are ignored, since they are present in both samples, and
only the residuals $\delta r'$ that are larger than 30\% of the maximum 
residual $\delta r_{\rm{max}}$ in a given cluster are considered. 
This has the effect of producing a 
narrowly-peaked distribution with a longer tail on the high end for tracks,
and a very broad, almost flat, distribution for showers. A possible
selection criterion for discriminating between them is to require $\delta r'$ to be
above the value where the electron distribution intersects the high-end tail of the muon
distribution. For 0.5\,GeV, the selection $\delta r' >4$\,cm will 
identify approximately 96\% of electrons as showers and only 6\% of muons as showers. 
At 1.5\,GeV, the intersection value increases to $\delta r' = $12\,cm, 
degrading the shower identification efficiency to 86\% for electrons, while slightly
improving the muon shower misidentification probability to 3\%. However, this selection
criteria is energy dependent, meaning that the energy of the cluster needs to be known
before a decision can be made as to whether the particle is a shower or a track.
To avoid this difficulty, a common selection value is imposed on all clusters
irrespective of their energy, namely that a shower must have
at least 90\% of its $\delta r'$ residuals to be longer than 2\,cm.
This gives a comparable 
performance to the energy-dependent selection for 0.5\,GeV particles, but offers no 
discrimination power at 1.5\,GeV or higher energies. This can be understood by looking
at the example events shown in Fig.~\ref{fig:emu1_5GeV}. For the muon, there are 
additional hit points that are perpendicular to the general direction of the track 
which originate from (delta) electrons that are knocked-off neighbouring atoms as 
the muon passes by. The residuals of these extra hits are large enough to be 
comparable to the typical residuals observed for electron showers. To remedy this problem, 
an additional variable is used, namely the ratio of the transverse-to-longitudinal 
extent $d_{\rm{T}}/d_{\rm{L}}$ of a 
convex hull volume that defines the outer edge which encompasses all of
the hits in the cluster~\cite{hull}. Here, the longitudinal component $d_{\rm{L}}$ 
is defined to be the hull length along the principal axis of the cluster, while the transverse 
component $d_{\rm{T}}$ is the sum of the two lengths that are orthogonal
to $d_{\rm{L}}$. In general, showers will have larger
convex hull ratios when compared to tracks. Figure~\ref{fig:convHull} shows the
distributions of this quantity for 0.5 and 1.5\,GeV electrons and muons after the previously
defined selections on the residuals $\delta r'$ have been applied. Track-shower
discrimination at high energies is restored, with minimal impact on the performance at
low energies, by requiring the convex hull ratio to be
larger than 0.12 for all cluster energies, which corresponds to tracks having a length
about 8 times longer than the transverse extent of any hit point filaments originating
from delta electrons.

\begin{table}[!htb]
\caption{Shower identification efficiencies for electron and muon 
monoenergetic particles based on 1,000-event samples}
\centering
\begin{tabular}{lll}
\hline
Generated & Electron & Muon \\
energy (GeV) & efficiency (\%) & efficiency (\%) \\
\hline
0.5 & $94.7 \pm 0.7$ & $3.8 \pm 0.6$ \\
1.0 & $98.4 \pm 0.4$ & $2.1 \pm 0.5$ \\
1.5 & $99.5 \pm 0.2$ & $1.9 \pm 0.4$ \\
2.0 & $99.8 \pm 0.1$ & $1.3 \pm 0.4$ \\
2.5 & $99.9 \pm 0.1$ & $1.4 \pm 0.4$ \\
3.0 & 100.0 & $0.9 \pm 0.3$ \\
\hline
\end{tabular}
\label{tab:shower}
\end{table}
Table~\ref{tab:shower} provides a summary of the shower identification efficiencies
for electrons and muons; a cluster is classified as a shower if at least 90\% of 
the $\delta r'$ residuals are larger than 2\,cm, and if it has a convex-hull ratio 
above 0.12. In fact, these selection requirements produce the optimal separation 
between tracks and showers based on the significance 
defined as $\epsilon_e/\sqrt{\epsilon_e + \epsilon_{\mu}}$, where 
$\epsilon_e$ ($\epsilon_{\mu}$) is the efficiency of identifying an electron 
(muon) as a shower. For the previous 770\,MeV neutrino to muon-proton event
sample discussed in Sect.~\ref{sec:mup}, the probability of misidentifying muons 
(protons) as showers is $6.6 \pm 0.6$\% ($7.9 \pm 0.6$\%), which is slightly worse
than the expected value of approximately 4\% owing to some of the original hits being
left out of the reconstructed clusters (see the efficiencies in 
Table~\ref{tab:mupeff}), which will affect the distributions of the residuals and 
convex-hull ratios.

\section{Electron--proton neutrino interactions: 
$\nu_e + n   \rightarrow e + p$}
\label{sec:ep}

\begin{figure}[hbt]
\subfigure[]{
  \includegraphics[width=0.45\textwidth]{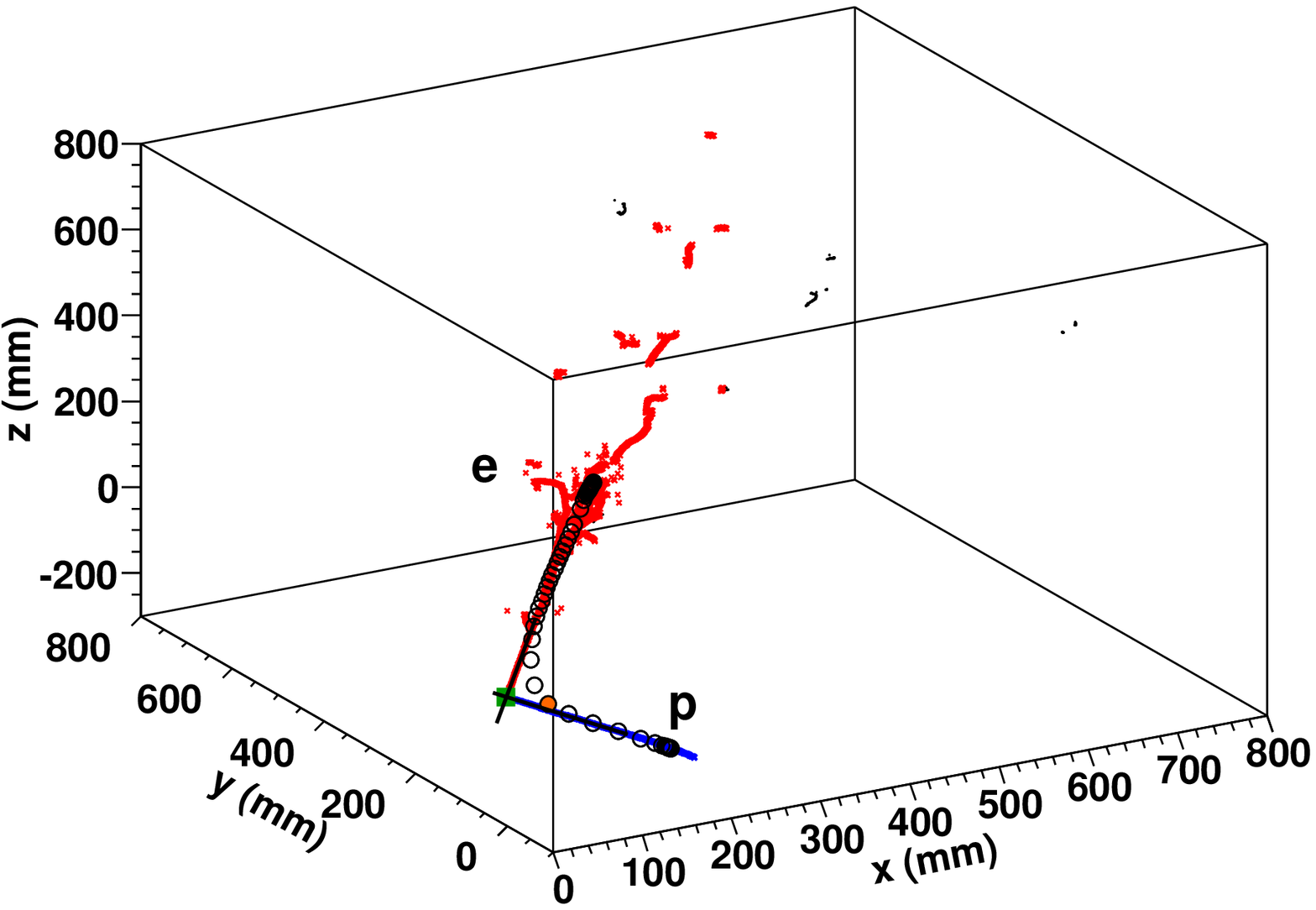}
}
\subfigure[]{
  \includegraphics[width=0.45\textwidth]{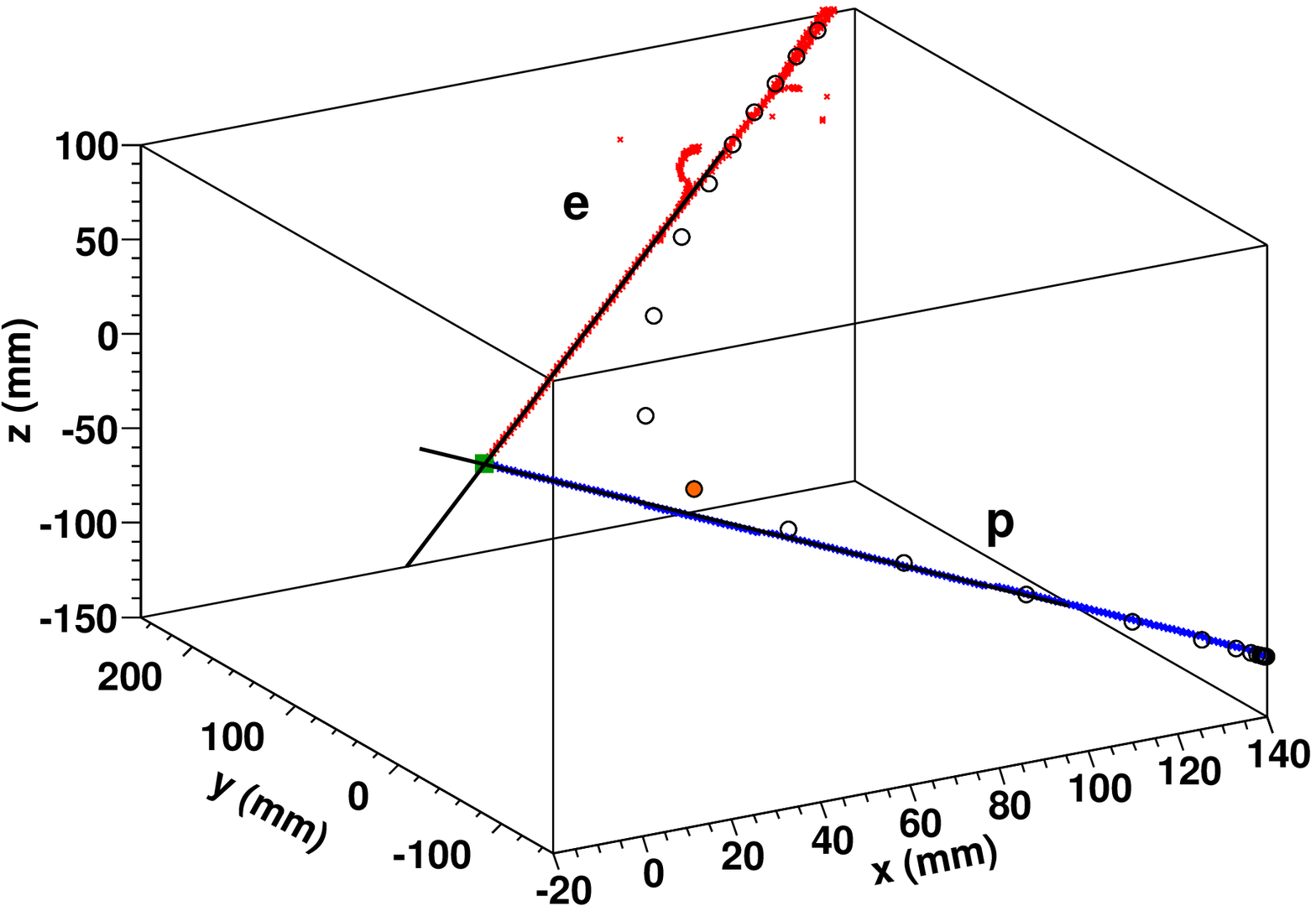}
}
\caption{An example lpc reconstruction of an electron-proton event showing
the hits associated to the electron shower (red) and proton track (blue) together
with the calculated lpc points (open circles). Also shown are the 
two line segments used to find the position of the primary 
interaction vertex (green square). These lines are made from
hits on either side of the feature point of the principal curve, which 
has the largest $1-|\rm{cos}\phi|$ value, and is shown as an 
orange-filled circle. Plot (b) is a 
close-up view of the interaction region of plot (a)}
\label{fig:epEvent}
\end{figure}

We now have all of the ingredients to fully reconstruct electron-proton events,
which have a two-prong topology involving a short proton track and an electron which
initially starts off like a track but quickly produces a cascade of hits in the
form of an electromagnetic shower, resulting in a halo of hits along the initial 
direction of the electron. Figure~\ref{fig:epEvent} shows an example 770\,MeV 
electron-neutrino interacting with a neutron to produce an electron shower and 
proton track originating from a common vertex point. 
The calculated points of the principal curve follow the 
hits of the proton track closely. They then bend around the vertex region to 
follow the hits in the initial track-like segment of the electron, then 
continue along the principal axis of the shower until the end of the core
region has been reached. The primary vertex is reconstructed using exactly the same
extended two-line method that was used for muon-proton events in Sect.~\ref{sec:mup}.
As before, the main feature point (with $1 - |\rm{cos}\phi| > 0.01$)
is used to find the extended straight line sections for the proton 
and track-like segment of the electron, and the reconstructed 
primary vertex corresponds to their point of closest approach. Then, two clusters
are formed, one on each side of the vertex, from hits that are closest to these 
straight lines. Further hits are added to each cluster by continuing 
along the principal curve direction initially given by the straight line section 
and adding hits that are closest to the remaining lpc points that have
$\delta r$ residuals below 10\,cm.

\begin{figure}[htb]
\subfigure[]{
  \includegraphics[width=0.45\textwidth]{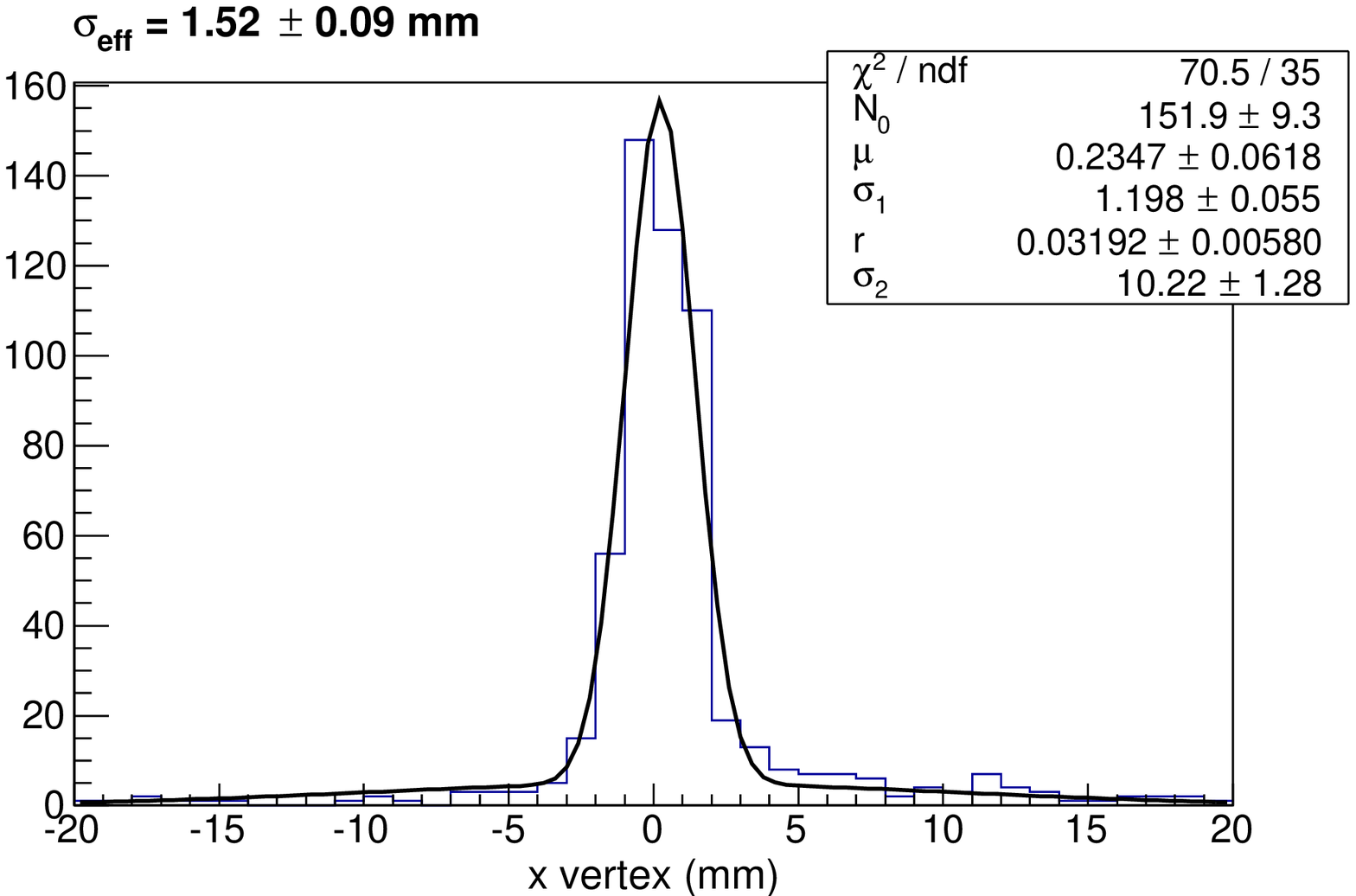}
}
\subfigure[]{
  \includegraphics[width=0.45\textwidth]{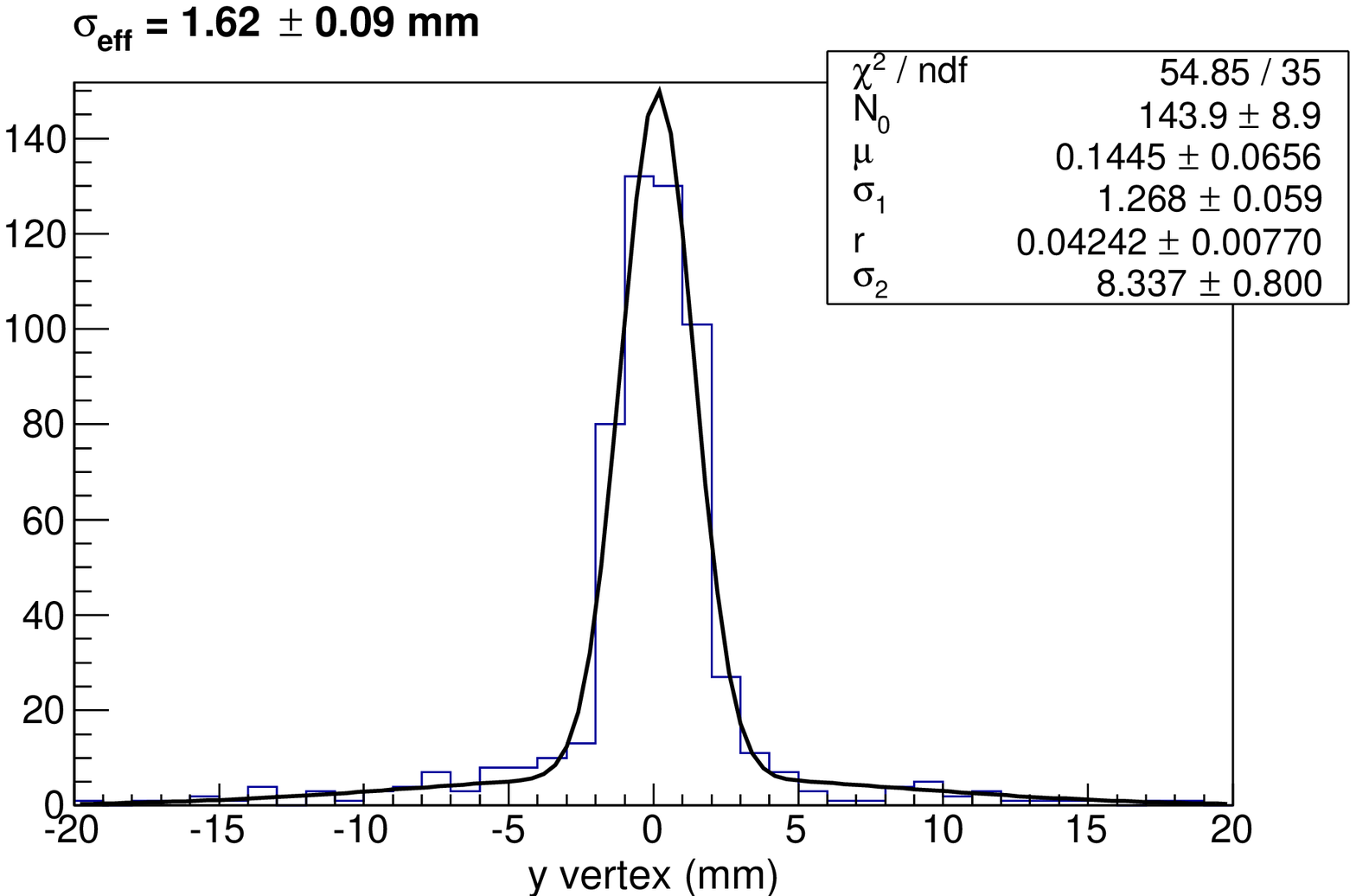}
}
\subfigure[]{
  \includegraphics[width=0.45\textwidth]{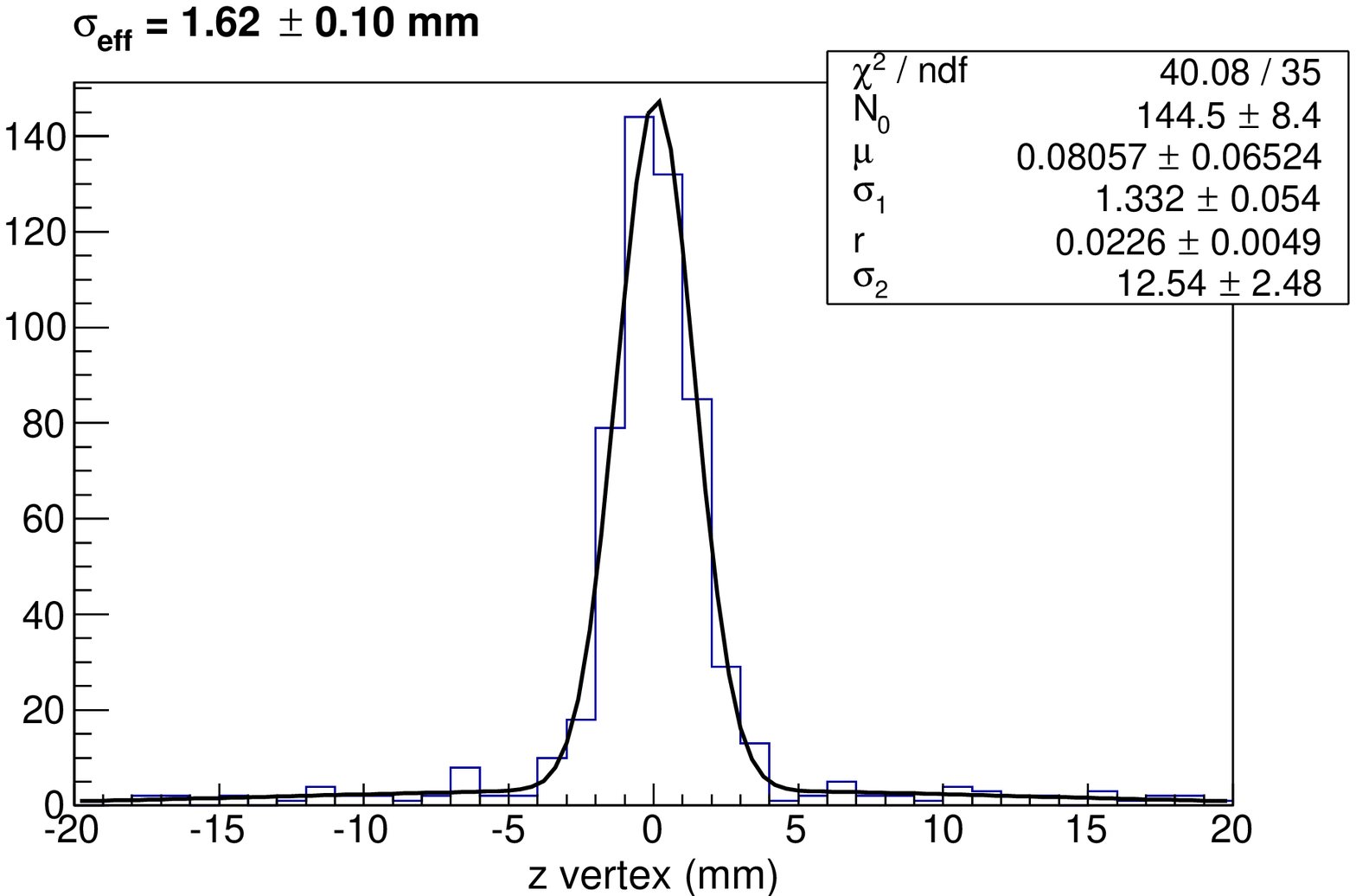}
}
\caption{Double Gaussian fits to the distributions of the 
reconstructed primary vertex position in $x$, $y$ and $z$ for electron-proton events.
Approximately two-thirds of events have a primary vertex found within 
2\,cm from the true vertex position (0,0,0)}
\label{fig:epvtx}
\end{figure}
\begin{figure}[htb]
\includegraphics[width=0.5\textwidth]{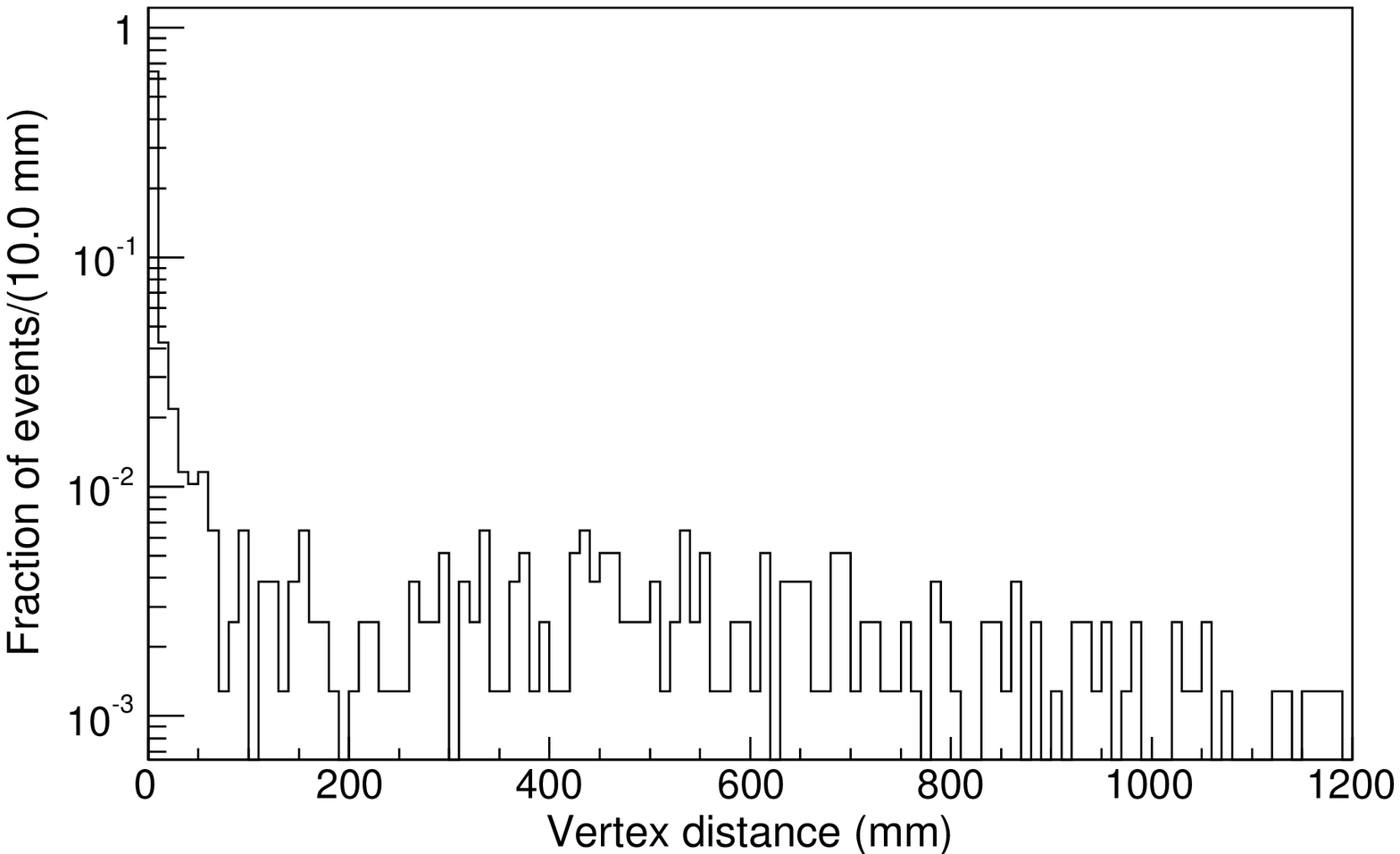}
\caption{The distribution of the distance of the 
reconstructed primary vertex position from the generated position (0,0,0) for
electron-proton events that pass the range selection on the proton track}
\label{fig:epVtxdist}
\end{figure}
\begin{figure}[htb]
\includegraphics[width=0.5\textwidth]{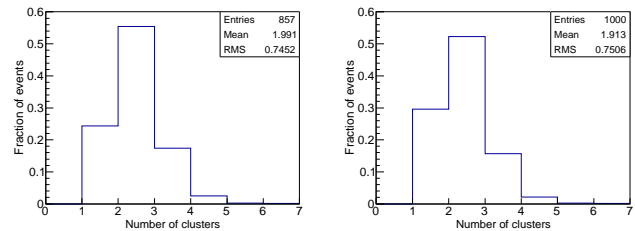}
\caption{The distribution of reconstructed clusters in a sample of 1000 electron-proton events
for (left) events passing a range selection on the proton track and (right) with
no range requirement imposed}
\label{fig:epNCl}
\end{figure}
Figure~\ref{fig:epvtx} shows double Gaussian fits to the primary vertex position
in $x$, $y$ and $z$ for a sample of 770\,MeV neutrino to electron-proton events
when the proton track has a minimum number of 25 hits. Approximately $15 \pm 1$\% of
protons in a sample of $\nu_e + n \rightarrow e + p$ events will not satisfy this
hit requirement. The vertexing resolution, taken to be the effective
width of the corresponding double Gaussian fit, is slightly worse than
the resolution obtained for muon-proton events. This is to be expected, since 
the onset of the electromagnetic shower will produce hits that will be some 
distance away from the initial direction of the electron, affecting the 
accuracy of finding the extended line sections. As was the case for
muon-proton events, the determination of the $x$ position of the vertex 
point has a slight positive bias of $0.23 \pm 0.06$\,mm. Despite the rather
good vertex resolution of approximately 1.5 to 1.6\,mm in each direction, only 67\% 
of the events that pass the proton hit requirement
have a vertex found within $\pm 2$\,cm from the generated position at $(0,0,0)$,
as illustrated in Fig.~\ref{fig:epVtxdist}.
Note that this 33\% inefficiency includes cases where no vertex is found.
This is significantly worse than the vertexing efficiency for muon-proton events (90\%)
due to the fact that hits from the shower can distract the
principal curve algorithm from picking up the proton track. Indeed, 
the random nature of the hit positions in the shower can induce multiple
feature points to be found along the principal curve. Figure~\ref{fig:epNCl}
shows the number of reconstructed clusters 
for the sample of electron-proton events with and without the proton
hit selection requirement. Most events have two clusters reconstructed, as expected,
although for about 25\% of the selected events, or 30\% in the full sample, 
the proton track has not been found. Approximately 20\% of the remaining 
events have more than two clusters found, meaning that a primary vertex 
(with the lowest $x$ co-ordinate) has been reconstructed together with 
secondary vertices embedded inside the electron shower. This is about a factor of
two higher than the number of multiple vertices found for muon-proton events,
despite including the requirement that neighbouring clusters are merged, and the 
vertex between them removed, if the angle between their principal axes 
is less than 20 degrees.

As was done for the muon-proton sample, the parameters of the 
principal curve defined in Table~\ref{tab:lpcparam} were optimised 
in order to give, on average, two clusters per electron-proton event, as well as 
providing maximal cluster and hit efficiencies and purities for the reconstructed 
electron and proton clusters. Again, the most important parameters are
the kernel width $h$, the step size $t$ and the number of lpc points $N_p$.
Table~\ref{tab:epeff} shows the results from this optimisation, including the
average efficiencies of identifying the electron and proton clusters as showers,
$93.4 \pm 0.6$\% and $11.9 \pm 0.7$\% respectively,
using the procedure described in Sect.~\ref{sec:shower}.
Note that the efficiency of reconstructing the proton cluster is strongly 
dependent on the efficiency of finding a primary vertex. It was found that 
variations to the step size within the range 0.04 to 0.06 did not significantly 
affect the reconstruction performance, provided the kernel-to-step size 
ratio was kept near values between 1.6 and 1.8. Additionally, using between 
100 and 300 lpc points produced similar results. 

\begin{table}[!bt]
\caption{Optimised performance of the lpc algorithm for 770\,MeV neutrino
to electron-proton events that satisfy the proton range requirement (857 out of an 
initial sample of 1000). Efficiencies and purities are averaged over all 
selected events}
\centering
\begin{tabular}{ll}
\hline
Quantity & Value \\
\hline
Lpc scaled kernel bandwidth $h$ & 0.072 \\
Lpc scaled step size $t$ & 0.040 \\
Number of lpc points & 100 \\
Fraction of events with no vertex found & $24.4 \pm 1.5$ \% \\
Electron cluster efficiency & $99.6 \pm 0.2$ \% \\
Electron hit efficiency & $74.5 \pm 1.5$ \% \\
Electron reconstruction efficiency & $74.3 \pm 1.5$ \% \\
Electron hit purity & $97.5 \pm 0.5$ \% \\
Electron shower efficiency & $93.4 \pm 0.6$ \% \\
Proton cluster efficiency & $63.9 \pm 1.6$ \% \\
Proton hit efficiency & $92.8 \pm 0.9$ \% \\
Proton reconstruction efficiency & $59.3 \pm 1.7$ \% \\
Proton hit purity & $96.8 \pm 0.6$ \% \\
Proton shower efficiency & $11.9 \pm 0.7$ \% \\
Vertex efficiency within $\pm 2$\,cm & $67.3 \pm 0.9$ \% \\
Vertex $x$ position resolution & $1.52 \pm 0.09$\,mm \\
Vertex $y$ position resolution & $1.62 \pm 0.09$\,mm \\
Vertex $z$ position resolution & $1.62 \pm 0.10$\,mm \\
\hline
\end{tabular}
\label{tab:epeff}
\end{table}
\begin{figure}[htb]
\subfigure[]{
  \includegraphics[width=0.45\textwidth]{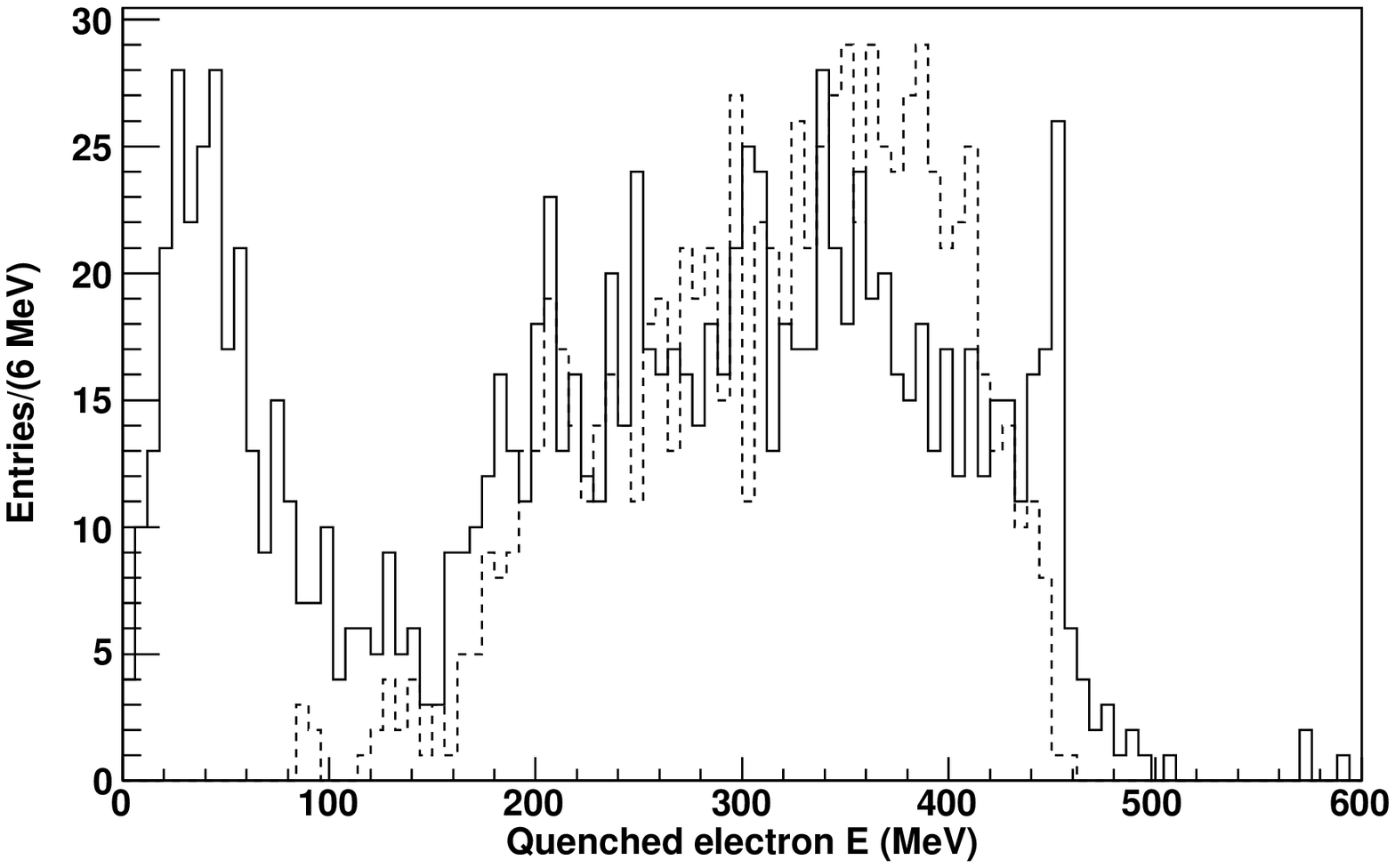}
}
\subfigure[]{
  \includegraphics[width=0.45\textwidth]{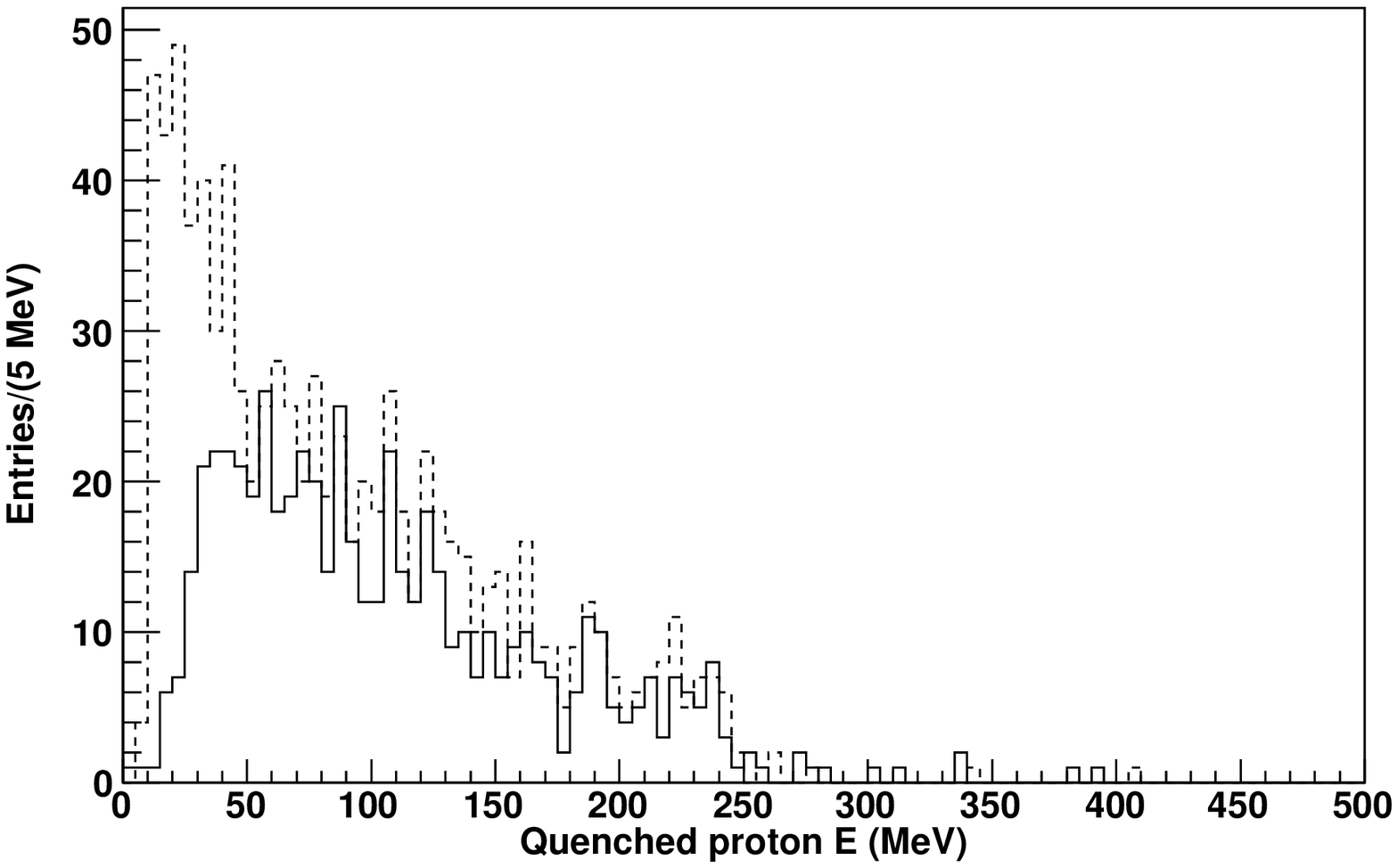}
}
\caption{Plots (a) and (b) show the reconstructed (solid line) and generated 
(dotted line) quenched energy histograms for the electrons and protons in 770\,MeV 
$\nu_e + n \rightarrow e + p$ events}
\label{fig:epEnergies}
\end{figure}
\begin{figure}[htb]
\subfigure[]{
  \includegraphics[width=0.45\textwidth]{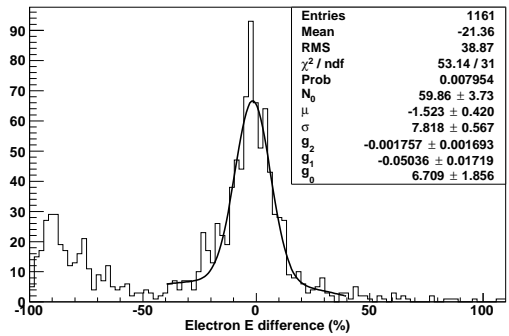}
}
\subfigure[]{
  \includegraphics[width=0.45\textwidth]{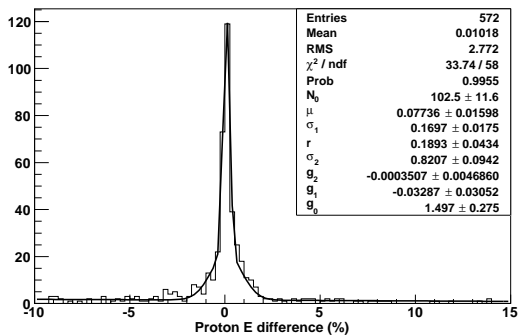}
}
\caption{Distributions of the fractional difference $f_E$ between the generated and 
reconstructed quenched energies for (a) electrons and (b) protons shown in 
Fig.~\ref{fig:epEnergies}. The main peak of the 
electron (proton) distribution is fitted with a single (double) Gaussian with a quadratic 
polynomial background $g_2f_E^2 + g_1f_E + g_0$. The height of each Gaussian function 
is given by $N_0$, while $\mu$ and $\sigma$ denote the mean and width, respectively. 
The relative normalisation between the two Gaussian terms (same mean, different 
widths $\sigma_1$ and $\sigma_2$) for the proton fit is given by the parameter $r$}
\label{fig:epEResolution}
\end{figure}
Despite most of the electrons being correctly found and
identified as showers, about 25\% of their hits are not included in the
(main) cluster. Some of these missing hits are at the outer edge of the shower hit cloud, with
$\delta r$ residuals larger than 10\,cm. However, most of them are misclassified as a separate
cluster within the main shower, which happens if more than one vertex has been reconstructed.
This will have a direct effect on the reconstructed energy of the electron cluster, which
is simply taken to be the sum of the energy deposits $Q_i$ of all of the associated hits.
Figure~\ref{fig:epEnergies} shows the generated quenched energy distributions for electrons 
and protons (770\,MeV $\nu_e$ events), as well as the reconstructed energies
of the electron and proton clusters when most of their hits
have the correct particle type. We can see that there is a
secondary peak below 100\,MeV for the electron distribution, corresponding to the 
missing hits in the main cluster. Furthermore, most of the 
protons that are not reconstructed have generated (quenched) energies below 50\,MeV;
the hits from these protons are instead associated to the main electron cluster, leading to
reconstructed energies that are higher than the generated values.
Figure~\ref{fig:epEResolution} shows the distributions of the fractional 
energy difference $f_E = (E_{\rm{reco}} - E_{\rm{gen}})/E_{\rm{gen}}$, where $E_{\rm{reco}}$ is the
reconstructed cluster energy and $E_{\rm{gen}}$ is the generated particle energy.
The $f_E$ distribution for electrons has a main single-Gaussian peak with a width
corresponding to a quenched energy resolution of approximately 8\%. The secondary
peak occurs for $f_E$ values below $-50$\%, and corresponds to additional clusters
found inside the electron shower. The $f_E$ distribution for correctly identified protons 
has a very narrow peak, which is to be expected since the proton hit efficiency and purity are 
both above 92\%. Fitting a double Gaussian function (common mean, two widths) to this peak gives an
effective quenched energy resolution approximately equal to 0.3\% for protons.
This compares well to the effective energy resolution of approximately 0.2\% for 
the reconstructed clusters in the previously mentioned muon-proton sample.

\begin{figure}[htb]
\subfigure[]{
  \includegraphics[width=0.4\textwidth]{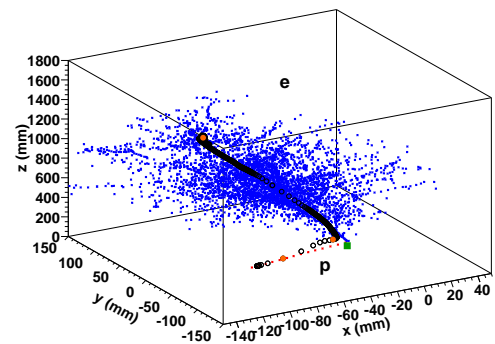}
}
\subfigure[]{
  \includegraphics[width=0.4\textwidth]{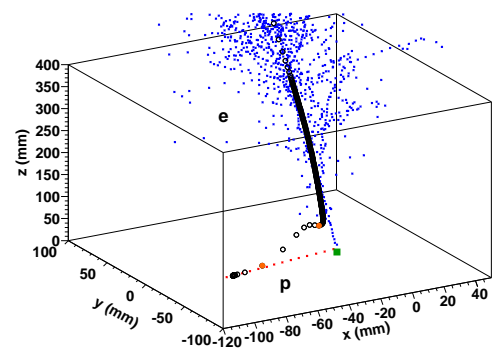}
}
\caption{The lpc reconstruction of a high energy electron-proton event showing
the hits associated to the 9.0\,GeV electron shower (blue) and 0.1\,GeV proton track 
(red) together with the calculated lpc points (open circles). Also shown are the 
feature points of the principal curve (orange-filled circles) and the reconstructed 
primary interaction vertex (green square). Plot (b) is a 
close-up view of the interaction region of plot (a)}
\label{fig:lagunaEvent}
\end{figure}

So far we have only looked at low-energy ($<1\,$GeV) neutrino interactions.
Figure~\ref{fig:lagunaEvent} shows the lpc reconstruction of a 
high energy ($\sim 9$\,GeV) electron-proton event. 
The electron shower extends over a very wide area, which means that the 
number of points in the principal curve needs to be significantly increased 
from around 100 to at least 500 in order to cover most of the core region of 
the shower, following the hits along the principal axis.
Additionally, the scaled kernel width needs to be increased
to the value 0.11 for a step size of 0.06. A close-up view of the event 
near the primary vertex region shows that the start of the electron shower is 
almost track-like, as was observed for the lower energy electrons. 
In this region, the principal curve 
starts to move away from the electron hits towards the proton track. 
During this transition,
the sheer number of hits in the shower start to push the principal curve back
towards itself. However, the local nature of the algorithm forces the points
back onto the proton track. This push-and-pull effect creates two feature points
which are actually close enough to create a merged range of lpc
points that are used for finding the primary vertex location via
the extended straight line method described in Sect.~\ref{sec:mup}. 
Inside the electron shower, there 
is an additional feature point near the end of the principal curve, owing to the
rather wide spread of hits at the edge of the shower affecting the convergence of the
curve, producing large angles between the remaining eigenvectors.

In most high-energy events, the algorithm finds multiple feature points
inside the core of the shower which adversely affects the performance of correctly finding
the primary vertex location, especially if the proton track has a 
very small range and is not well separated from the shower. In order
to consistently reconstruct high-energy events correctly, other tools
and methods need to be developed and incorporated into the lpc algorithm.

\section{Conclusion}
\label{sec:con}

We have presented a local principal curve algorithm that can 
reconstruct neutrino interaction events in liquid argon. 
The algorithm creates a series of (three-dimensional) points that follows 
the local density of hits.
It does so by calculating the localised mean shift, which changes
direction based on the largest eigenvector obtained from the $3 \times 3$ covariance matrix
of a set of weights which determine the size and shape of the local neighbourhood 
of points. Differences in the angle $\phi$ between consecutive eigenvectors can produce
peaks in the $1 - |\rm{cos}\phi|$ distribution when the principal curve is rapidly
changing direction. These peaks correspond to feature points which 
highlight the presence of interaction vertices, which can be reconstructed by 
finding the point of closest approach between two straight line sections associated 
to nearby hits on either side of a given feature point. Clusters can then be formed
by continuing along the principal curve direction initially 
given by each straight line section and adding hits that are closest to the
remaining lpc points. The residual distance between hits and their nearest lpc point,
together with a measure of the size of a convex hull encompassing all of the hits 
in a cluster, can be used to discriminate showers from tracks.
The reconstruction performance of the algorithm with regards to vertexing,
clustering, energy resolution and track-shower identification has been 
tested on 770\,MeV neutrino interaction muon-proton (CCQE) and electron-proton 
events. For high-energy events, further work is required to better use the
information provided by the increased number of feature points.

There are possible further uses of this algorithm. For example, it
is straightforward to use it to identify clusters when the hit positions
are only known in two dimensions; the third co-ordinate for all hits is just
set to zero or ignored, and all other procedures remain the same.
Additionally, it should be possible to use the algorithm
to find feature points and reconstruct clusters for events when more than two particles
originate from a common vertex point. Here, points along the principal curve 
will only be able to follow the two main particles which will contain the majority 
of the hits. To find the other particles, the hits associated to the two reconstructed 
clusters need to be removed and the algorithm re-run on the remaining hits.
However, care must be taken to avoid removing hits unnecessarily within
such a procedure. A possible way to proceed is to use the ratio between the first and
second largest eigenvalues of the covariance matrix defined in Eq.~\ref{eqn:sigma}, which 
may indicate the presence of bifurcation (``branching'') points that can act as 
starting locations for further principal curves. At the time of writing, we have 
not yet studied a strategy for reconstructing multi-particle final state interactions.

To conclude, the local principal curve algorithm provides a wealth of information that
can be used to automatically reconstruct neutrino interaction events in liquid 
argon detectors.

\begin{acknowledgements}
We thank Daniel Brunt, Harmanjeet Khera and Jamie Wynn
who all worked with the group as part of the
University of Warwick project student programme. This work was
financially supported by the Science and Technology Facilities Council 
consolidated grant ST/H00369X/1 and by the European Community under the
European Commission Framework Programme 7 grant 
LAGUNA-LBNO/FP7-INFRASTRUCTURES/284518.
\end{acknowledgements}

\end{document}